
%
%
%
%
%
%
%
%
\def\standardrisposta{s }\def\reducedrisposta{r }
\def\mplarisposta{mpla }\def\zerorisposta{z }
\def\doublerisposta{d }\def\cartarisposta{e }\def\amsrisposta{y }
\newcount\ingrandimento \newcount\sinnota \newcount\dimnota
\newcount\unoduecol \newdimen\collhsize \newdimen\tothsize
\newdimen\fullhsize \newcount\controllorisposta \sinnota=1
\newskip\infralinea  \global\controllorisposta=0
\immediate\write16 { ********  Welcome to PANDA macros (Plain TeX,
AP, 1991) ******** }
\immediate\write16 { You'll have to answer a few questions in
lowercase.}
\message{>  Do you want it in double-page (d), reduced (r)
or standard format (s) ? }\read-1 to\risposta
\message{>  Do you want it in USA A4 (u) or EUROPEAN A4
(e) paper size ? }\read-1 to\srisposta
\message{>  Do you have AMSFonts 2.0 (math) fonts (y/n) ? }
\read-1 to\arisposta
%
%
%
%
%
\ifx\risposta\standardrisposta \ingrandimento=1200
\message {>> This will come out UNREDUCED << }
\dimnota=2 \unoduecol=1 \global\controllorisposta=1 \fi
\ifx\risposta\reducedrisposta \ingrandimento=1095 \dimnota=1
\unoduecol=1  \global\controllorisposta=1
\message {>> This will come out REDUCED << } \fi
\ifx\risposta\doublerisposta \ingrandimento=1000 \dimnota=2
\unoduecol=2   \message {>> You must print this in
LANDSCAPE orientation << } \global\controllorisposta=1 \fi
\ifx\risposta\mplarisposta \ingrandimento=1000 \dimnota=1
\message {>> Mod. Phys. Lett. A format << }
\unoduecol=1 \global\controllorisposta=1 \fi
\ifx\risposta\zerorisposta \ingrandimento=1000 \dimnota=2
\message {>> Zero Magnification format << }
\unoduecol=1 \global\controllorisposta=1 \fi
\ifnum\controllorisposta=0  \ingrandimento=1200
\message {>>> ERROR IN INPUT, I ASSUME STANDARD
UNREDUCED FORMAT <<< }  \dimnota=2 \unoduecol=1 \fi
\magnification=\ingrandimento
%
%
%
%
\newdimen\eucolumnsize \newdimen\eudoublehsize \newdimen\eudoublevsize
\newdimen\uscolumnsize \newdimen\usdoublehsize \newdimen\usdoublevsize
\newdimen\eusinglehsize \newdimen\eusinglevsize \newdimen\ussinglehsize
\newskip\standardbaselineskip \newdimen\ussinglevsize
\newskip\reducedbaselineskip \newskip\doublebaselineskip
\eucolumnsize=12.0truecm    
\eudoublehsize=25.5truecm   
\eudoublevsize=6.5truein    
\uscolumnsize=4.4truein     
\usdoublehsize=9.4truein    
\usdoublevsize=6.8truein    
\eusinglehsize=6.5truein    
\eusinglevsize=24truecm     
\ussinglehsize=6.5truein    
\ussinglevsize=8.9truein    
\standardbaselineskip=16pt plus.2pt  
\reducedbaselineskip=14pt plus.2pt   
\doublebaselineskip=12pt plus.2pt    
%
%
\def\Portoffset{}
\def\Landoffset{}
\ifx\risposta\mplarisposta \def\Portoffset{\hoffset=1.8truecm} \fi
%
%
\def\Landspec{}
\tolerance=10000
\parskip=0pt plus2pt  \leftskip=0pt \rightskip=0pt
%
%
\ifx\risposta\standardrisposta \infralinea=\standardbaselineskip \fi
\ifx\risposta\reducedrisposta  \infralinea=\reducedbaselineskip \fi
\ifx\risposta\doublerisposta   \infralinea=\doublebaselineskip \fi
\ifx\risposta\mplarisposta     \infralinea=13pt \fi
\ifx\risposta\zerorisposta     \infralinea=12pt plus.2pt\fi
\ifnum\controllorisposta=0    \infralinea=\standardbaselineskip \fi
\ifx\risposta\doublerisposta   \Landoffset \else \Portoffset \fi
\ifx\risposta\doublerisposta \ifx\srisposta\cartarisposta
\tothsize=\eudoublehsize \collhsize=\eucolumnsize
\vsize=\eudoublevsize  \else  \tothsize=\usdoublehsize
\collhsize=\uscolumnsize \vsize=\usdoublevsize \fi \else
\ifx\srisposta\cartarisposta \tothsize=\eusinglehsize
\vsize=\eusinglevsize \else  \tothsize=\ussinglehsize
\vsize=\ussinglevsize \fi \collhsize=4.4truein \fi
\ifx\risposta\mplarisposta \tothsize=5.0truein
\vsize=7.8truein \collhsize=4.4truein \fi
%
%
%
%
\newcount\contaeuler \newcount\contacyrill \newcount\contaams
\font\ninerm=cmr9  \font\eightrm=cmr8  \font\sixrm=cmr6
\font\ninei=cmmi9  \font\eighti=cmmi8  \font\sixi=cmmi6
\font\ninesy=cmsy9  \font\eightsy=cmsy8  \font\sixsy=cmsy6
\font\ninebf=cmbx9  \font\eightbf=cmbx8  \font\sixbf=cmbx6
\font\ninett=cmtt9  \font\eighttt=cmtt8  \font\nineit=cmti9
\font\eightit=cmti8 \font\ninesl=cmsl9  \font\eightsl=cmsl8
\skewchar\ninei='177 \skewchar\eighti='177 \skewchar\sixi='177
\skewchar\ninesy='60 \skewchar\eightsy='60 \skewchar\sixsy='60
\hyphenchar\ninett=-1 \hyphenchar\eighttt=-1 \hyphenchar\tentt=-1
%
\font\tencmmib=cmmib10  \newfam\cmmibfam  \skewchar\tencmmib='177
\font\tencmbsy=cmbsy10  \newfam\cmbsyfam  \skewchar\tencmbsy='60
\def\scaps{\cmcsc}                 
\font\tencmcsc=cmcsc10  \newfam\cmcscfam
\ifnum\ingrandimento=1095

\font\capsone=cmcsc10 at 10.95pt 

\else

\font\capsone=cmcsc10 at 12pt 
\fi

\def\ttaarr{\bf}		
\def\ppaarr{\sl}		

%
%
%
\newfam\eufmfam \newfam\msamfam \newfam\msbmfam \newfam\eufbfam
\def\Loadeulerfonts{\global\contaeuler=1 \ifx\arisposta\amsrisposta
\font\teneufm=eufm10              
\font\eighteufm=eufm8 \font\nineeufm=eufm9 \font\sixeufm=eufm6
\font\seveneufm=eufm7  \font\fiveeufm=eufm5
\font\teneufb=eufb10              
\font\eighteufb=eufb8 \font\nineeufb=eufb9 \font\sixeufb=eufb6
\font\seveneufb=eufb7  \font\fiveeufb=eufb5
\font\teneurm=eurm10              
\font\eighteurm=eurm8 \font\nineeurm=eurm9
\font\teneurb=eurb10              
\font\eighteurb=eurb8 \font\nineeurb=eurb9
\font\teneusm=eusm10              
\font\eighteusm=eusm8 \font\nineeusm=eusm9
\font\teneusb=eusb10              
\font\eighteusb=eusb8 \font\nineeusb=eusb9
\else \def\eufm{\tt} \def\eufb{\tt} \def\eurm{\tt} \def\eurb{\tt}
\def\eusm{\tt} \def\eusb{\tt}    \fi}
\def\loadeuler{\Loadeulerfonts\tenpoint}
\def\loadamsmath{\global\contaams=1 \ifx\arisposta\amsrisposta
\font\tenmsam=msam10 \font\ninemsam=msam9 \font\eightmsam=msam8
\font\sevenmsam=msam7 \font\sixmsam=msam6 \font\fivemsam=msam5
\font\tenmsbm=msbm10 \font\ninemsbm=msbm9 \font\eightmsbm=msbm8
\font\sevenmsbm=msbm7 \font\sixmsbm=msbm6 \font\fivemsbm=msbm5
\else \def\msbm{\bf} \fi \def\Bbb{\msbm} \def\symbl{\msam} \tenpoint}
\def\loadcyrill{\global\contacyrill=1 \ifx\arisposta\amsrisposta
\font\tenwncyr=wncyr10 \font\ninewncyr=wncyr9 \font\eightwncyr=wncyr8
\font\tenwncyb=wncyr10 \font\ninewncyb=wncyr9 \font\eightwncyb=wncyr8
\font\tenwncyi=wncyr10 \font\ninewncyi=wncyr9 \font\eightwncyi=wncyr8
\else \def\cyrill{\sl} \def\cyrilb{\sl} \def\cyrili{\sl} \fi\tenpoint}
\ifx\arisposta\amsrisposta
\font\sevenex=cmex7               
\font\eightex=cmex8  \font\nineex=cmex9
\font\ninecmmib=cmmib9   \font\eightcmmib=cmmib8
\font\sevencmmib=cmmib7 \font\sixcmmib=cmmib6
\font\fivecmmib=cmmib5   \skewchar\ninecmmib='177
\skewchar\eightcmmib='177  \skewchar\sevencmmib='177
\skewchar\sixcmmib='177   \skewchar\fivecmmib='177
\font\ninecmbsy=cmbsy9    \font\eightcmbsy=cmbsy8
\font\sevencmbsy=cmbsy7  \font\sixcmbsy=cmbsy6
\font\fivecmbsy=cmbsy5   \skewchar\ninecmbsy='60
\skewchar\eightcmbsy='60  \skewchar\sevencmbsy='60
\skewchar\sixcmbsy='60    \skewchar\fivecmbsy='60
\font\ninecmcsc=cmcsc9    \font\eightcmcsc=cmcsc8     \else
\def\cmmib{\fam\cmmibfam\tencmmib}\textfont\cmmibfam=\tencmmib
\scriptfont\cmmibfam=\tencmmib \scriptscriptfont\cmmibfam=\tencmmib
\def\cmbsy{\fam\cmbsyfam\tencmbsy} \textfont\cmbsyfam=\tencmbsy
\scriptfont\cmbsyfam=\tencmbsy \scriptscriptfont\cmbsyfam=\tencmbsy
\scriptfont\cmcscfam=\tencmcsc \scriptscriptfont\cmcscfam=\tencmcsc
\def\cmcsc{\fam\cmcscfam\tencmcsc} \textfont\cmcscfam=\tencmcsc \fi
\catcode`@=11
\newskip\ttglue
\gdef\tenpoint{\def\rm{\fam0\tenrm}
  \textfont0=\tenrm \scriptfont0=\sevenrm \scriptscriptfont0=\fiverm
  \textfont1=\teni \scriptfont1=\seveni \scriptscriptfont1=\fivei
  \textfont2=\tensy \scriptfont2=\sevensy \scriptscriptfont2=\fivesy
  \textfont3=\tenex \scriptfont3=\tenex \scriptscriptfont3=\tenex
  \def\mcal{\fam2 \tensy}  \def\mmit{\fam1 \teni}
  \textfont\itfam=\tenit \def\it{\fam\itfam\tenit}
  \textfont\slfam=\tensl \def\sl{\fam\slfam\tensl}
  \textfont\ttfam=\tentt \scriptfont\ttfam=\eighttt
  \scriptscriptfont\ttfam=\eighttt  \def\tt{\fam\ttfam\tentt}
  \textfont\bffam=\tenbf \scriptfont\bffam=\sevenbf
  \scriptscriptfont\bffam=\fivebf \def\bf{\fam\bffam\tenbf}
     \ifx\arisposta\amsrisposta    \ifnum\contaeuler=1
  \textfont\eufmfam=\teneufm \scriptfont\eufmfam=\seveneufm
  \scriptscriptfont\eufmfam=\fiveeufm \def\eufm{\fam\eufmfam\teneufm}
  \textfont\eufbfam=\teneufb \scriptfont\eufbfam=\seveneufb
  \scriptscriptfont\eufbfam=\fiveeufb \def\eufb{\fam\eufbfam\teneufb}
  \def\eurm{\teneurm} \def\eurb{\teneurb} \def\eusm{\teneusm}
  \def\eusb{\teneusb}    \fi    \ifnum\contaams=1
  \textfont\msamfam=\tenmsam \scriptfont\msamfam=\sevenmsam
  \scriptscriptfont\msamfam=\fivemsam \def\msam{\fam\msamfam\tenmsam}
  \textfont\msbmfam=\tenmsbm \scriptfont\msbmfam=\sevenmsbm
  \scriptscriptfont\msbmfam=\fivemsbm \def\msbm{\fam\msbmfam\tenmsbm}
     \fi      \ifnum\contacyrill=1     \def\cyrill{\tenwncyr}
  \def\cyrilb{\tenwncyb}  \def\cyrili{\tenwncyi}         \fi
  \textfont3=\tenex \scriptfont3=\sevenex \scriptscriptfont3=\sevenex
  \def\cmmib{\fam\cmmibfam\tencmmib} \scriptfont\cmmibfam=\sevencmmib
  \textfont\cmmibfam=\tencmmib  \scriptscriptfont\cmmibfam=\fivecmmib
  \def\cmbsy{\fam\cmbsyfam\tencmbsy} \scriptfont\cmbsyfam=\sevencmbsy
  \textfont\cmbsyfam=\tencmbsy  \scriptscriptfont\cmbsyfam=\fivecmbsy
  \def\cmcsc{\fam\cmcscfam\tencmcsc} \scriptfont\cmcscfam=\eightcmcsc
  \textfont\cmcscfam=\tencmcsc \scriptscriptfont\cmcscfam=\eightcmcsc
     \fi            \tt \ttglue=.5em plus.25em minus.15em
  \normalbaselineskip=12pt
  \setbox\strutbox=\hbox{\vrule height8.5pt depth3.5pt width0pt}
  \let\sc=\eightrm \let\big=\tenbig   \normalbaselines
  \baselineskip=\infralinea  \rm}
\gdef\ninepoint{\def\rm{\fam0\ninerm}
  \textfont0=\ninerm \scriptfont0=\sixrm \scriptscriptfont0=\fiverm
  \textfont1=\ninei \scriptfont1=\sixi \scriptscriptfont1=\fivei
  \textfont2=\ninesy \scriptfont2=\sixsy \scriptscriptfont2=\fivesy
  \textfont3=\tenex \scriptfont3=\tenex \scriptscriptfont3=\tenex
  \def\mcal{\fam2 \ninesy}  \def\mmit{\fam1 \ninei}
  \textfont\itfam=\nineit \def\it{\fam\itfam\nineit}
  \textfont\slfam=\ninesl \def\sl{\fam\slfam\ninesl}
  \textfont\ttfam=\ninett \scriptfont\ttfam=\eighttt
  \scriptscriptfont\ttfam=\eighttt \def\tt{\fam\ttfam\ninett}
  \textfont\bffam=\ninebf \scriptfont\bffam=\sixbf
  \scriptscriptfont\bffam=\fivebf \def\bf{\fam\bffam\ninebf}
     \ifx\arisposta\amsrisposta  \ifnum\contaeuler=1
  \textfont\eufmfam=\nineeufm \scriptfont\eufmfam=\sixeufm
  \scriptscriptfont\eufmfam=\fiveeufm \def\eufm{\fam\eufmfam\nineeufm}
  \textfont\eufbfam=\nineeufb \scriptfont\eufbfam=\sixeufb
  \scriptscriptfont\eufbfam=\fiveeufb \def\eufb{\fam\eufbfam\nineeufb}
  \def\eurm{\nineeurm} \def\eurb{\nineeurb} \def\eusm{\nineeusm}
  \def\eusb{\nineeusb}     \fi   \ifnum\contaams=1
  \textfont\msamfam=\ninemsam \scriptfont\msamfam=\sixmsam
  \scriptscriptfont\msamfam=\fivemsam \def\msam{\fam\msamfam\ninemsam}
  \textfont\msbmfam=\ninemsbm \scriptfont\msbmfam=\sixmsbm
  \scriptscriptfont\msbmfam=\fivemsbm \def\msbm{\fam\msbmfam\ninemsbm}
     \fi       \ifnum\contacyrill=1     \def\cyrill{\ninewncyr}
  \def\cyrilb{\ninewncyb}  \def\cyrili{\ninewncyi}         \fi
  \textfont3=\nineex \scriptfont3=\sevenex \scriptscriptfont3=\sevenex
  \def\cmmib{\fam\cmmibfam\ninecmmib}  \textfont\cmmibfam=\ninecmmib
  \scriptfont\cmmibfam=\sixcmmib \scriptscriptfont\cmmibfam=\fivecmmib
  \def\cmbsy{\fam\cmbsyfam\ninecmbsy}  \textfont\cmbsyfam=\ninecmbsy
  \scriptfont\cmbsyfam=\sixcmbsy \scriptscriptfont\cmbsyfam=\fivecmbsy
  \def\cmcsc{\fam\cmcscfam\ninecmcsc} \scriptfont\cmcscfam=\eightcmcsc
  \textfont\cmcscfam=\ninecmcsc \scriptscriptfont\cmcscfam=\eightcmcsc
     \fi            \tt \ttglue=.5em plus.25em minus.15em
  \normalbaselineskip=11pt
  \setbox\strutbox=\hbox{\vrule height8pt depth3pt width0pt}
  \let\sc=\sevenrm \let\big=\ninebig \normalbaselines\rm}
\gdef\eightpoint{\def\rm{\fam0\eightrm}
  \textfont0=\eightrm \scriptfont0=\sixrm \scriptscriptfont0=\fiverm
  \textfont1=\eighti \scriptfont1=\sixi \scriptscriptfont1=\fivei
  \textfont2=\eightsy \scriptfont2=\sixsy \scriptscriptfont2=\fivesy
  \textfont3=\tenex \scriptfont3=\tenex \scriptscriptfont3=\tenex
  \def\mcal{\fam2 \eightsy}  \def\mmit{\fam1 \eighti}
  \textfont\itfam=\eightit \def\it{\fam\itfam\eightit}
  \textfont\slfam=\eightsl \def\sl{\fam\slfam\eightsl}
  \textfont\ttfam=\eighttt \scriptfont\ttfam=\eighttt
  \scriptscriptfont\ttfam=\eighttt \def\tt{\fam\ttfam\eighttt}
  \textfont\bffam=\eightbf \scriptfont\bffam=\sixbf
  \scriptscriptfont\bffam=\fivebf \def\bf{\fam\bffam\eightbf}
     \ifx\arisposta\amsrisposta   \ifnum\contaeuler=1
  \textfont\eufmfam=\eighteufm \scriptfont\eufmfam=\sixeufm
  \scriptscriptfont\eufmfam=\fiveeufm \def\eufm{\fam\eufmfam\eighteufm}
  \textfont\eufbfam=\eighteufb \scriptfont\eufbfam=\sixeufb
  \scriptscriptfont\eufbfam=\fiveeufb \def\eufb{\fam\eufbfam\eighteufb}
  \def\eurm{\eighteurm} \def\eurb{\eighteurb} \def\eusm{\eighteusm}
  \def\eusb{\eighteusb}       \fi    \ifnum\contaams=1
  \textfont\msamfam=\eightmsam \scriptfont\msamfam=\sixmsam
  \scriptscriptfont\msamfam=\fivemsam \def\msam{\fam\msamfam\eightmsam}
  \textfont\msbmfam=\eightmsbm \scriptfont\msbmfam=\sixmsbm
  \scriptscriptfont\msbmfam=\fivemsbm \def\msbm{\fam\msbmfam\eightmsbm}
     \fi       \ifnum\contacyrill=1     \def\cyrill{\eightwncyr}
  \def\cyrilb{\eightwncyb}  \def\cyrili{\eightwncyi}         \fi
  \textfont3=\eightex \scriptfont3=\sevenex \scriptscriptfont3=\sevenex
  \def\cmmib{\fam\cmmibfam\eightcmmib}  \textfont\cmmibfam=\eightcmmib
  \scriptfont\cmmibfam=\sixcmmib \scriptscriptfont\cmmibfam=\fivecmmib
  \def\cmbsy{\fam\cmbsyfam\eightcmbsy}  \textfont\cmbsyfam=\eightcmbsy
  \scriptfont\cmbsyfam=\sixcmbsy \scriptscriptfont\cmbsyfam=\fivecmbsy
  \def\cmcsc{\fam\cmcscfam\eightcmcsc} \scriptfont\cmcscfam=\eightcmcsc
  \textfont\cmcscfam=\eightcmcsc \scriptscriptfont\cmcscfam=\eightcmcsc
     \fi             \tt \ttglue=.5em plus.25em minus.15em
  \normalbaselineskip=9pt
  \setbox\strutbox=\hbox{\vrule height7pt depth2pt width0pt}
  \let\sc=\sixrm \let\big=\eightbig \normalbaselines\rm }
\gdef\tenbig#1{{\hbox{$\left#1\vbox to8.5pt{}\right.\n@space$}}}
\gdef\ninebig#1{{\hbox{$\textfont0=\tenrm\textfont2=\tensy
   \left#1\vbox to7.25pt{}\right.\n@space$}}}
\gdef\eightbig#1{{\hbox{$\textfont0=\ninerm\textfont2=\ninesy
   \left#1\vbox to6.5pt{}\right.\n@space$}}}
\def\alternativefont#1#2{\ifx\arisposta\amsrisposta \relax \else
\xdef#1{#2} \fi}
\global\contaeuler=0 \global\contacyrill=0 \global\contaams=0
%
%
%
%
\newbox\fotlinebb \newbox\hedlinebb \newbox\leftcolumn
\gdef\makeheadline{\vbox to 0pt{\vskip-22.5pt
     \fullline{\vbox to8.5pt{}\the\headline}\vss}\nointerlineskip}
\gdef\makehedlinebb{\vbox to 0pt{\vskip-22.5pt
     \fullline{\vbox to8.5pt{}\copy\hedlinebb\hfil
     \line{\hfill\the\headline\hfill}}\vss} \nointerlineskip}
\gdef\makefootline{\baselineskip=24pt \fullline{\the\footline}}
\gdef\makefotlinebb{\baselineskip=24pt
    \fullline{\copy\fotlinebb\hfil\line{\hfill\the\footline\hfill}}}
\gdef\doubleformat{\shipout\vbox{\Landspec\makehedlinebb
     \fullline{\box\leftcolumn\hfil\columnbox}\makefotlinebb}
     \advancepageno}
\gdef\columnbox{\leftline{\pagebody}}
\gdef\line#1{\hbox to\hsize{\hskip\leftskip#1\hskip\rightskip}}
\gdef\fullline#1{\hbox to\fullhsize{\hskip\leftskip{#1}%
\hskip\rightskip}}
\gdef\footnote#1{\let\@sf=\empty
         \ifhmode\edef\#sf{\spacefactor=\the\spacefactor}\/\fi
         #1\@sf\vfootnote{#1}}
\gdef\vfootnote#1{\insert\footins\bgroup
         \ifnum\dimnota=1  \eightpoint\fi
         \ifnum\dimnota=2  \ninepoint\fi
         \ifnum\dimnota=0  \tenpoint\fi
         \interlinepenalty=\interfootnotelinepenalty
         \splittopskip=\ht\strutbox
         \splitmaxdepth=\dp\strutbox \floatingpenalty=20000
         \leftskip=\oldssposta \rightskip=\olddsposta
         \spaceskip=0pt \xspaceskip=0pt
         \ifnum\sinnota=0   \textindent{#1}\fi
         \ifnum\sinnota=1   \item{#1}\fi
         \footstrut\futurelet\next\fo@t}
\gdef\fo@t{\ifcat\bgroup\noexpand\next \let\next\f@@t
             \else\let\next\f@t\fi \next}
\gdef\f@@t{\bgroup\aftergroup\@foot\let\next}
\gdef\f@t#1{#1\@foot} \gdef\@foot{\strut\egroup}
\gdef\footstrut{\vbox to\splittopskip{}}
\skip\footins=\bigskipamount
\count\footins=1000  \dimen\footins=8in
\catcode`@=12
\tenpoint
\ifnum\unoduecol=1 \hsize=\tothsize   \fullhsize=\tothsize \fi
\ifnum\unoduecol=2 \hsize=\collhsize  \fullhsize=\tothsize \fi
\global\let\lrcol=L      \ifnum\unoduecol=1
\output{\plainoutput{\ifnum\tipbnota=2 \clearnmbnota\fi}} \fi
\ifnum\unoduecol=2 \output{\if L\lrcol
     \global\setbox\leftcolumn=\columnbox
     \global\setbox\fotlinebb=\line{\hfill\the\footline\hfill}
     \global\setbox\hedlinebb=\line{\hfill\the\headline\hfill}
     \advancepageno  \global\let\lrcol=R
     \else  \doubleformat \global\let\lrcol=L \fi
     \ifnum\outputpenalty>-20000 \else\dosupereject\fi
     \ifnum\tipbnota=2\clearnmbnota\fi }\fi
\def\ifdoublepage{\ifnum\unoduecol=2 }
\gdef\yespagenumbers{\footline={\hss\tenrm\folio\hss}}
\gdef\ciao{ \ifnum\fdefcontre=1 \endfdef\fi
     \par\vfill\supereject \ifnum\unoduecol=2
     \if R\lrcol  \headline={}\nopagenumbers\null\vfill\eject
     \fi\fi \end}

\newskip\olddsposta \newskip\oldssposta
\global\oldssposta=\leftskip \global\olddsposta=\rightskip

\def\filldots{\leaders\hbox to 1em{\hss.\hss}\hfill}
\def\inquadrb#1 {\vbox {\hrule  \hbox{\vrule \vbox {\vskip .2cm
    \hbox {\ #1\ } \vskip .2cm } \vrule  }  \hrule} }
 \def\newline{\hfil\break}
\def\jump{\vskip\baselineskip} \newskip\iinnffrr
\def\sjump{\iinnffrr=\baselineskip
          \divide\iinnffrr by 2 \vskip\iinnffrr}
\def\bjump{\vskip\baselineskip \vskip\baselineskip}
\newcount\nmbnota  \def\clearnmbnota{\global\nmbnota=0}
\newcount\tipbnota \def\letterfootnote{\global\tipbnota=1}

\def\note#1{\global\advance\nmbnota by 1 \ifnum\tipbnota=1
    \footnote{$^{\rm\nttlett}$}{#1} \else {\ifnum\tipbnota=2
    \footnote{$^{\nttsymb}$}{#1}
    \else\footnote{$^{\the\nmbnota}$}{#1}\fi}\fi}
\def\nttlett{\ifcase\nmbnota \or a\or b\or c\or d\or e\or f\or
g\or h\or i\or j\or k\or l\or m\or n\or o\or p\or q\or r\or
s\or t\or u\or v\or w\or y\or x\or z\fi}
\def\nttsymb{\ifcase\nmbnota \or\dag\or\sharp\or\ddag\or\star\or
\natural\or\flat\or\clubsuit\or\diamondsuit\or\heartsuit
\or\spadesuit\fi}   \clearnmbnota
\def\numberfootnote{\global\tipbnota=0} \numberfootnote
\def\setnote#1{\expandafter\xdef\csname#1\endcsname{
\ifnum\tipbnota=1 {\rm\nttlett} \else {\ifnum\tipbnota=2
{\nttsymb} \else \the\nmbnota\fi}\fi} }
\newcount\nbmfig  \def\clearnbmfig{\global\nbmfig=0}
\gdef\figure{\global\advance\nbmfig by 1
      {\rm fig. \the\nbmfig}}   \clearnbmfig
\def\setfig#1{\expandafter\xdef\csname#1\endcsname{fig. \the\nbmfig}}
 \def\endformula{\eqno\numero $$}
 \def\efr{\endformula}
\newcount\frmcount \def\clearfrmcount{\global\frmcount=0}
\def\numero{\global\advance\frmcount by 1   \ifnum\indappcount=0
  {\ifnum\cpcount <1 {\hbox{\rm (\the\frmcount )}}  \else
  {\hbox{\rm (\the\cpcount .\the\frmcount )}} \fi}  \else
  {\hbox{\rm (\applett .\the\frmcount )}} \fi}
\def\nameformula#1{\global\advance\frmcount by 1%
\ifnum\draftnum=0  {\ifnum\indappcount=0%
{\ifnum\cpcount<1\xdef\spzzttrra{(\the\frmcount )}%
\else\xdef\spzzttrra{(\the\cpcount .\the\frmcount )}\fi}%
\else\xdef\spzzttrra{(\applett .\the\frmcount )}\fi}%
\else\xdef\spzzttrra{(#1)}\fi%
\expandafter\xdef\csname#1\endcsname{\spzzttrra}
\eqno \hbox{\rm\spzzttrra} $$}
\def\nfr{\nameformula}    
\def\nameali#1{\global\advance\frmcount by 1%
\ifnum\draftnum=0  {\ifnum\indappcount=0%
{\ifnum\cpcount<1\xdef\spzzttrra{(\the\frmcount )}%
\else\xdef\spzzttrra{(\the\cpcount .\the\frmcount )}\fi}%
\else\xdef\spzzttrra{(\applett .\the\frmcount )}\fi}%
\else\xdef\spzzttrra{(#1)}\fi%
\expandafter\xdef\csname#1\endcsname{\spzzttrra}
  \hbox{\rm\spzzttrra} }      \clearfrmcount
\newcount\cpcount \def\clearcpcount{\global\cpcount=0}
\newcount\subcpcount \def\clearsubcpcount{\global\subcpcount=0}
\newcount\appcount \def\clearappcount{\global\appcount=0}
\newcount\indappcount \def\clearindappcount{\indappcount=0}
\newcount\sottoparcount 

\def\applett{\ifcase\appcount  \or {A}\or {B}\or {C}\or
{D}\or {E}\or {F}\or {G}\or {H}\or {I}\or {J}\or {K}\or {L}\or
{M}\or {N}\or {O}\or {P}\or {Q}\or {R}\or {S}\or {T}\or {U}\or
{V}\or {W}\or {X}\or {Y}\or {Z}\fi    \ifnum\appcount<0
\immediate\write16 {Panda ERROR - Appendix: counter "appcount"
out of range}\fi  \ifnum\appcount>26  \immediate\write16 {Panda
ERROR - Appendix: counter "appcount" out of range}\fi}
\clearappcount  \clearindappcount \newcount\connttrre
\def\clearconnttrre{\global\connttrre=0} \newcount\countref
\def\clearcountref{\global\countref=0} \clearcountref
\def\chapter#1{\global\advance\cpcount by 1 \clearfrmcount
                 \goodbreak\null\vbox{\jump\nobreak
                 \clearsubcpcount\clearindappcount
                 \itemitem{\ttaarr\the\cpcount .\qquad}{\ttaarr #1}
                 \par\nobreak\jump\sjump}\nobreak}
\def\section#1{\global\advance\subcpcount by 1 \goodbreak\null
               \vbox{\sjump\nobreak\ifnum\indappcount=0
                 {\ifnum\cpcount=0 {\itemitem{\ppaarr
               .\the\subcpcount\quad\enskip\ }{\ppaarr #1}\par} \else
                 {\itemitem{\ppaarr\the\cpcount .\the\subcpcount\quad
                  \enskip\ }{\ppaarr #1} \par}  \fi}
                \else{\itemitem{\ppaarr\applett .\the\subcpcount\quad
                 \enskip\ }{\ppaarr #1}\par}\fi\nobreak\jump}\nobreak}
\clearsubcpcount
\def\appendix#1{\global\advance\appcount by 1 \clearfrmcount
                  \goodbreak\null\vbox{\jump\nobreak
                  \global\advance\indappcount by 1 \clearsubcpcount
          \itemitem{ }{\hskip-40pt\ttaarr Appendix\ \applett :\ #1}
             \nobreak\jump\sjump}\nobreak}
\clearappcount \clearindappcount
\def\references{\goodbreak\null\vbox{\jump\nobreak
   \itemitem{}{\ttaarr References} \nobreak\jump\sjump}\nobreak}

\clearcpcount\clearcountref

\def\setchap#1{\ifnum\indappcount=0{\ifnum\subcpcount=0%
\xdef\spzzttrra{\the\cpcount}%
\else\xdef\spzzttrra{\the\cpcount .\the\subcpcount}\fi}
\else{\ifnum\subcpcount=0 \xdef\spzzttrra{\applett}%
\else\xdef\spzzttrra{\applett .\the\subcpcount}\fi}\fi
\expandafter\xdef\csname#1\endcsname{\spzzttrra}}
\newcount\draftnum \newcount\ppora   \newcount\ppminuti
\global\ppora=\time   \global\ppminuti=\time
\global\divide\ppora by 60  \draftnum=\ppora
\multiply\draftnum by 60    \global\advance\ppminuti by -\draftnum
\def\droggi{\number\day /\number\month /\number\year\ \the\ppora
:\the\ppminuti}     \global\draftnum=0
\def\draftcomment#1{\ifnum\draftnum=0 \relax \else
{\ {\bf ***}\ #1\ {\bf ***}\ }\fi} 
%
%
\catcode`@=11
\gdef\Ref#1{\expandafter\ifx\csname @rrxx@#1\endcsname\relax%
{\global\advance\countref by 1    \ifnum\countref>200
\immediate\write16 {Panda ERROR - Ref: maximum number of references
exceeded}  \expandafter\xdef\csname @rrxx@#1\endcsname{0}\else
\expandafter\xdef\csname @rrxx@#1\endcsname{\the\countref}\fi}\fi
\ifnum\draftnum=0 \csname @rrxx@#1\endcsname \else#1\fi}
\gdef\beginref{\ifnum\draftnum=0  \gdef\Rref{\fairef}
\gdef\endref{\scriviref} \else\relax\fi
\ifx\risposta\mplarisposta \ninepoint \fi
\parskip 2pt plus.2pt \baselineskip=12pt}
\def\Reflab#1{[#1]} \gdef\Rref#1#2{\item{\Reflab{#1}}{#2}}
\gdef\endref{\relax}  \newcount\conttemp
\gdef\fairef#1#2{\expandafter\ifx\csname @rrxx@#1\endcsname\relax
{\global\conttemp=0 \immediate\write16 {Panda ERROR - Ref: reference
[#1] undefined}} \else
{\global\conttemp=\csname @rrxx@#1\endcsname } \fi
\global\advance\conttemp by 50  \global\setbox\conttemp=\hbox{#2} }
\gdef\scriviref{\clearconnttrre\conttemp=50
\loop\ifnum\connttrre<\countref \advance\conttemp by 1
\advance\connttrre by 1
\item{\Reflab{\the\connttrre}}{\unhcopy\conttemp} \repeat}
\clearcountref \clearconnttrre
\catcode`@=12
\ifx\risposta\mplarisposta \def\Reflab#1{#1.} \letterfootnote \fi

\def\slashchar#1{\setbox0=\hbox{$#1$} \dimen0=\wd0
     \setbox1=\hbox{/} \dimen1=\wd1 \ifdim\dimen0>\dimen1
      \rlap{\hbox to \dimen0{\hfil/\hfil}} #1 \else
      \rlap{\hbox to \dimen1{\hfil$#1$\hfil}} / \fi}
\ifx\oldchi\undefined \let\oldchi=\chi
  \def\cchi{{\raise 1pt\hbox{$\oldchi$}}} \let\chi=\cchi \fi
\def\square{\hbox{{$\sqcup$}\llap{$\sqcap$}}}

\def\frac#1#2{{\textstyle{#1 \over #2}}}

\def\half{\ifinner {\scriptstyle {1 \over 2}}\else {1 \over 2} \fi}

\def\simge{\rlap{\raise 2pt \hbox{$>$}}{\lower 2pt \hbox{$\sim$}}}
\def\simle{\rlap{\raise 2pt \hbox{$<$}}{\lower 2pt \hbox{$\sim$}}}

\def\vbig#1#2{{\vbigd@men=#2\divide\vbigd@men by 2%
\hbox{$\left#1\vbox to \vbigd@men{}\right.\n@space$}}}

%
%
\newcount\fdefcontre \newcount\fdefcount \newcount\indcount
\newread\filefdef  \newread\fileftmp  \newwrite\filefdef
\newwrite\fileftmp     \def\strip#1*.A {#1}
\def\futuredef#1{\beginfdef
\expandafter\ifx\csname#1\endcsname\relax%
{\immediate\write\fileftmp {#1*.A}
\immediate\write16 {Panda Warning - fdef: macro "#1" on page
\the\pageno \space undefined}
\ifnum\draftnum=0 \expandafter\xdef\csname#1\endcsname{(?)}
\else \expandafter\xdef\csname#1\endcsname{(#1)} \fi
\global\advance\fdefcount by 1}\fi   \csname#1\endcsname}

\def\beginfdef{\ifnum\fdefcontre=0
\immediate\openin\filefdef \jobname.fdef
\immediate\openout\fileftmp \jobname.ftmp
\global\fdefcontre=1  \ifeof\filefdef \immediate\write16 {Panda
WARNING - fdef: file \jobname.fdef not found, run TeX again}
\else \immediate\read\filefdef to\spzzttrra
\global\advance\fdefcount by \spzzttrra
\indcount=0      \loop\ifnum\indcount<\fdefcount
\advance\indcount by 1   \immediate\read\filefdef to\spezttrra
\immediate\read\filefdef to\sppzttrra
\edef\spzzttrra{\expandafter\strip\spezttrra}
\immediate\write\fileftmp {\spzzttrra *.A}
\expandafter\xdef\csname\spzzttrra\endcsname{\sppzttrra}
\repeat \fi \immediate\closein\filefdef \fi}
\def\endfdef{\immediate\closeout\fileftmp   \ifnum\fdefcount>0
\immediate\openin\fileftmp \jobname.ftmp
\immediate\openout\filefdef \jobname.fdef
\immediate\write\filefdef {\the\fdefcount}   \indcount=0
\loop\ifnum\indcount<\fdefcount    \advance\indcount by 1
\immediate\read\fileftmp to\spezttrra
\edef\spzzttrra{\expandafter\strip\spezttrra}
\immediate\write\filefdef{\spzzttrra *.A}
\edef\spezttrra{\string{\csname\spzzttrra\endcsname\string}}
\iwritel\filefdef{\spezttrra}
\repeat  \immediate\closein\fileftmp \immediate\closeout\filefdef
\immediate\write16 {Panda Warning - fdef: Label(s) may have changed,
re-run TeX to get them right}\fi}
\def\iwritel#1#2{\newlinechar=-1
{\newlinechar=`\ \immediate\write#1{#2}}\newlinechar=-1}
\global\fdefcontre=0 \global\fdefcount=0 \global\indcount=0
%
%
\null
%
%
%
%
\loadamsmath
\loadeuler
%
%
%
%
\def\W{$\cal W$}
\def\ie{{\it i.e.\/}}

\def\gg{{\eufm g}}
\def\hh{{\eufm h}}
\def\ss{{\eufm s}}
\def\dd{{\eufm d}}
\def\s{{\bf s}}
\def\Vir{{\eufm Vir}}
\def\semdir{\mathrel\triangleright\joinrel\mathrel<}
\def\Pos{{\rm P}_{\geq0[\s]}}
\def\Neg{{\rm P}_{<0[\s]}}
\pageno=0
\nopagenumbers{\baselineskip=12pt
\line{\hfill CERN-TH.6998/93}
\line{\hfill US-FT-8/93}
\line{\hfill SWAT/92-93/10}
\line{\hfill\tt hep-th/9311067}
\line{\hfill October 1993}
\ifdoublepage \bjump\bjump\else\jump\vfill\fi
\centerline{\capsone ADDITIONAL SYMMETRIES OF GENERALIZED}
\sjump\sjump
\centerline{\capsone INTEGRABLE HIERARCHIES}
\jump\sjump
\centerline{
{\scaps Timothy J. Hollowood\footnote{$^1$}{{\tt
hollow@surya11.cern.ch}\newline
On leave from: Department of Mathematics, University College of Swansea,
Swansea SA2 8PP, U.K.}},
{\scaps J. Luis Miramontes\footnote{$^2$}{{\tt
miramont@cernvm.cern.ch}\newline Address after $1^{\rm st}$ October
1993: Departamento de F\'\i sica de Part\'\i culas,\newline
Universidad de Santiago, E-15706  Santiago de Compostela, Spain}}}
\sjump
\centerline{\sl Theory Division, CERN,}
\centerline{\sl CH-1211 Geneva 23, Switzerland}
\sjump  \centerline{and} \sjump
\centerline{\scaps Joaqu\'\i n S\'anchez Guill\'en\footnote{$^3$}{\tt
joaquin@egaes1.usc.es}}\sjump\sjump
\centerline{\sl Departamento de F\'\i sica de Part\'\i culas,}
\centerline{\sl Universidad de Santiago,}
\centerline{\sl E-15706 Santiago de Compostela, Spain}
\bjump
\ifdoublepage
\vfill
{\noindent
\line{CERN-TH.6998/93\hfill}
\line{October 1993\hfill}}
\eject\null\vfill\fi
\centerline{\capsone ABSTRACT}\sjump
The non-isospectral symmetries of a general class of integrable
hierarchies are
found, generalizing the Galilean and scaling symmetries
of the Korteweg--de Vries equation and its hierarchy.
The symmetries arise in a very natural way from the
semi-direct product structure of the
Virasoro algebra and the affine Kac--Moody algebra underlying the construction
of the hierarchy. In particular,
the generators of the symmetries are shown to satisfy a subalgebra of the
Virasoro algebra. When a tau-function formalism is available, the infinitesimal
symmetries act directly on the tau-functions as moments of Virasoro currents.
Some comments are made regarding the r\^ole of the non-isospectral symmetries
and the form of the string equations in matrix-model formulations of quantum
gravity in two-dimensions and related systems.
\sjump\vfill
\ifdoublepage \else
\noindent
\line{CERN-TH.6998/93\hfill}
\line{October 1993\hfill}\fi
\eject}
\yespagenumbers\pageno=1
\footline={\hss\tenrm-- \folio\ --\hss}
%
%
\chapter{Introduction.}

Since the discovery of the soliton solution of the
Korteweg--de Vries (KdV) equation
$$
{\partial u\over \partial t} = {1\over4} {\partial^3 u\over \partial x^3}
+ {3\over2} u{\partial u\over \partial x},
\nfr{KdV}
much effort has been devoted to elucidating the nature of the integrability of
soliton hierarchies (see~[\Ref{NEW}]
for a nice discussion of the history of the theory of soliton equations,
and~[\Ref{DAS}] for an introduction to integrable models). One of the
more important and surprising developments has been the recognition
of the deep connection existing between
integrable hierarchies of non-linear differential equations and infinite
dimensional Lie algebras. This connection has been manifested in two
apparently unconnected approaches.

In the ``tau-function'' approach, pioneered by the Japanese school~[\Ref{HIR},
\Ref{JAPOS}], the
equations are cast in a particular bilinear or Hirota form by the use
of a special set of variables---the tau-functions.
For instance, the original
variable and the tau-function of the KdV equation are related---in
standard conventions---by the well-known formula
$$
u\>=\> 2\> {\partial^2 \over \partial x^2}\, \ln \tau\,.
\nfr{TauKdV}
It was clear in the original work of~[\Ref{JAPOS}]
that the affine  Kac--Moody algebras play a central r\^ole in this
approach, but it was made even clearer by Kac and
Wakimoto~[\Ref{KWI}]. In this last work, the  authors construct
integrable hierarchies of equations in Hirota form associated to
vertex operator representations of affine Kac--Moody algebras.

The other approach is inspired by the seminal work of Drinfel'd and
Sokolov~[\Ref{DS}]. In this construction, the central objects are gauge
fields in the loop-algebra of a finite Lie algebra and the equations are the
conditions of zero-curvature on these gauge fields. In the original work
of~[\Ref{DS}], the
authors make use of the ``principal''  gradation of the loop-algebra
in an essential way; in particular, the  construction involves the
``principal Heisenberg subalgebra''. On the  other hand, it is well
known that the affine Kac--Moody algebras have other inequivalent Heisenberg
subalgebras~[\Ref{KBOOK},\Ref{KPET}], an observation that was
exploited in~[\Ref{GEN1}] (see also~[\Ref{GEN2}]) to construct more
general integrable hierarchies. In fact, it was shown in [\Ref{GEN3}] that
it is not necessary to restrict to the loop algebra and the constructions
of the hierarchies are completely representation independent; in particular,
they are
independent of the center. Moreover, when the affine Kac--Moody algebra
has a vertex operator representation
the zero-curvature hierarchies can be written in terms of
tau-functions~[\Ref{GEN3}] and hence a bridge can be established
between the two approaches.

The purpose of this paper is to complete the description of the
zero-curvature integrable hierarchies of~[\Ref{GEN1}] by discussing
their ``additional'' or ``non-isospectral'' symmetries\note{A
symmetry is called either ``isospectral'' or ``non-isospectral''
according to whether it
preserves or changes the spectrum of the auxiliary linear problem
associated to the non-linear equation,
respectively~[\Ref{LA},\Ref{GRIN}].}. In this context, a
``symmetry'' is some transformation that relates one solution of an equation
with another solution of the same
equation. Recall that, for example, the KdV equation has an
infinite set of commuting symmetries infinitesimally generated by the
infinite set of conserved hamiltonians. Hence, because the existence
of an infinite set of conserved hamiltonians is a necessary condition for
integrability, any integrable non-linear differential equation has
an infinite set of commuting symmetries; these are actually the
``isospectral''  symmetries. The corresponding infinite set of
infinitesimal generators are nothing else but the
commuting flows defining the ``integrable hierarchy'' associated to
the original non-linear differential equation. In the case of the
KdV equation, the general form of these flows is
$$ {\partial
u\over\partial t_{2k+1}} = P_{k}\left(u, \partial_x u, \partial_{x}^2
u, \ldots\right)\,, \nfr{Flow}
where $t_1\equiv x$, $t_3\equiv t$, and $P_k$ is a
polynomial function of $u$ and its $x$-derivatives; the original KdV
equation corresponding to  $k=1$. For each integer $k>1$, \Flow\
generates a symmetry of the KdV equation in  the sense that if $u$ is
a solution, so is $u\, +\,  \epsilon\,\partial u/\partial t_{2k+1}$,
$\epsilon\ll1$.

However, the group of symmetries of the KdV equation is
known to be much bigger still, since there exist the
Galilean and scaling transformations,
$$
\eqalign{
& u(x,t)\>\mapsto\> \widetilde{u}(x,t)\> =\> u(\,x + vt\,,\, t\,)
\> + \>{2\over3}\, v\,,  \cr
& u(x,t)\>\mapsto\> \widetilde{u}(x,t)\> =\> {\rm e}^{2r}\, u(\,{\rm
e}^{r} x\,,\,  {\rm e}^{3r} t\,)\,,\cr}
\nfr{Galilean}
respectively, for arbitrary values of $v$ and $r$. When $v$ and $r$ are
infinitesimal, one can write them as
$$
\eqalignno{
&\widetilde{u}(x,t) - u(x,t)\> \approx \> {2v\over3}\left(\, {3\over2}\, t\,
{\partial u \over\partial x}\> +\> 1\right)\/,&\nameali{Galinf}\cr
&\widetilde{u}(x,t) - u(x,t) \>\approx \> r\left(\, x\,{\partial u
\over\partial x}\> + \>
3\, t\,  {\partial u \over\partial t}\> +\> 2\, u \right)\/.&
\nameali{Scalinf}\cr}
$$
Notice that, in contrast to the flows defined by
\Flow, the  right-hand sides of \Galinf\ and \Scalinf\ involve
explicitly $x$ and $t$,
in addition to $u$ and its $x$-derivatives (hence the flows cannot be
hamiltonian). \Galinf\ and \Scalinf\
are just the first two generators of the infinite set of
``non-isospectral'' or ``additional'' symmetries of the KdV equation.
These  additional symmetries commute  with the isospectral ones, that is
the original flows of the KdV hierarchy, but do not commute among
themselves. Instead,
their generators satisfy part of the Virasoro algebra; the two
transformations in
\Galinf\ and \Scalinf\ corresponding to the $L_{-1}$ and the $L_0$
generators, respectively.
The connection of the additional symmetries and the Virasoro algebra can be
seen most clearly at the level of the tau-functions as discussed
in~[\Ref{VOS}], where it has been  shown that  they are
generated by the infinitesimal transformations of the  tau-function
$$ \tau \>\mapsto \>\tilde\tau \>=\> \tau \,+ \, \epsilon\, L_{m}
\tau\,, \nfr{Koos}
with $m\geq-1$, and $L_m$ being the generators of the Virasoro algebra
acting in the appropriate Fock space.

The additional (non-isospectral) symmetries of the KdV equation were first
found in~[\Ref{IBRA}]. Similar symmetries have also been found
in the  Kadomtsev-Petviashvili (KP) equation and its
hierarchy~[\Ref{CHEN}] and
in the Ablowitz-Kaup-Newell-Segur (AKNS) system~[\Ref{CHINOS}],
leading to
a general construction of non-isospectral symmetries of integrable
hierarchies by the inverse scattering method
in~[\Ref{ORLOV}] (see~[\Ref{GRIN}] for a more complete list of
references). In all these cases, the generators of the additional symmetries
satisfy a subalgebra of the
Virasoro algebra suggesting that there
is a natural action of (part of) the Virasoro algebra, or ${\rm Diff}(S^1)$,
on integrable hierarchies.

More recently, there has been a
renewed interest in the integrable hierarchies of partial differential
equations and their additional (non-isospectral) symmetries because of the
important r\^ole they play in the matrix-model formulation of
two-dimensional quantum and topological gravity~(see~[\Ref{REV}]
and references therein). In the continuum or {\it double-scaling\/} limit
the partition function of the matrix-models
is a tau-function of an integrable hierarchy\note{This also
seems to be true, in some cases, before taking the double-scaling limit.}.
Furthermore, this
tau-function is constrained by an additional equation known as the
``string equation'', which turns out to be the condition that the
relevant solution of the hierarchy is actually invariant under the additional
symmetries~[\Ref{LA}]. As has been said previously, the
generators of the additional symmetries satisfy a subalgebra of the
Virasoro algebra, and this is the reason why the
string equation appears as a set of Virasoro
constraints on the matrix-model partition function~[\Ref{VER},\Ref{FUK}].
On a slightly different tack, the Virasoro constraints also encode
the contact terms of physical
operators~[\Ref{WITT}] in some two-dimensional topological field theories.
These issues have motivated an investigation of the additional
symmetries of the Drinfel'd-Sokolov $A_n$-KdV hierarchies~[\Ref{LA},
\Ref{SEMI}], and those of
the KP hierarchy in~[\Ref{SEMI},\Ref{DICK}].

In this paper, we shall
discuss the additional symmetries of the zero-curvature hierarchies
of~[\Ref{GEN1}-\Ref{GEN3}] (some preliminary results have been presented
in~[\Ref{KIEV}]). It turns out that the
origin of the additional symmetries and of the Virasoro action on
the hierarchy is very natural. It is induced precisely by the
semi-direct product of the Virasoro algebra with the affine Kac--Moody
algebra in terms of which the hierarchy is defined. Furthermore,
both the flows of the zero-curvature hierarchies and the generators
of their additional symmetries are constructed in terms of the affine
Kac--Moody algebra in a completely representation independent way. It
is also shown that the action of the symmetries on the tau-functions
is precisely that of \Koos. In order to make the paper reasonably
self-contained, we summarize in section 2 the main results
of~[\Ref{GEN1}-\Ref{GEN3}]. The expression for the generators of the
additional symmetries are found in section 3 and they are shown to
satisfy a subalgebra of the  Virasoro algebra. Section 4 connects the results
to the tau-function formalism, recovering the results of~[\Ref{VOS}].
Finally, in section 5, we propose a
generalization of the ``string equation'' that selects the solution
that is invariant under the additional symmetries. This can be taken
as the starting point of an investigation into the possible relation between
the generalized
hierarchies and two-dimensional physical systems including quantum
gravity. Our conventions and some properties of
affine Kac--Moody algebras are presented in the appendix.

\chapter{The generalized zero-curvature hierarchies.}

In this section, we summarize the main results of~[\Ref{GEN1},\Ref{GEN3}],
to which one should refer for further
details. In order that the exposition be uninterrupted, we have
collected our conventions and notations concerning
affine Kac--Moody algebras in the appendix.

In~[\Ref{GEN1}], a generalized integrable hierarchy was associated
to each untwisted affine Kac--Moody algebra $\gg$, a particular Heisenberg
subalgebra $\ss\subset\gg$ (whose associated gradation is $\s'$), and an
additional gradation $\s$, such that $\s\preceq\s'$ with respect to a
partial ordering (see the appendix).

There is a flow of the hierarchy for each element of $\ss$ of
non-negative $\s'$-grade, this is the set $\{b_j\/,\ j\in E\geq0\}$.
The flows are defined in terms of ``Lax operators'', of the form
$$
{\cal L}_j = {\partial\over\partial t_j} - b_j - q(j)\/,\qquad
j\in E\geq0\/,
\nfr{Lax}
where $q(j)$ is a function of the $t_j$'s on the intersection
$$
Q(j) = \gg_{\geq0}(\s)\cap \gg_{<j}(\s')\/.
\nfr{Potential}
In order to ensure that the flows are uniquely associated to elements of
the set $\{b_j\,,\ j\in E\geq0\}$, we will demand that $q(j)$ has no
constant terms proportional to $b_i$ with $i<j$. The integrable hierarchy
of equations is defined by the zero curvature conditions
$$
[{\cal L}_i,{\cal L}_j]=0\/.
\nfr{Zerocurv}

In general, the above system of equations exhibits a {\it gauge
invariance} of the form
$$
{\cal L}_j\rightarrow U{\cal L}_j U^{-1}
\nfr{Gaugetrans}
preserving $q(j)\in Q(j)$, where $U$ is a function on the group generated
by the finite dimensional subalgebra given by the intersection
$$
P=\gg_0(\s) \cap \gg_{<0}(\s')\/.
\nfr{Gaugegroup}
The equations of the hierarchy are to be thought of as equations on the
equivalence of classes of $Q(j)$ under these gauge transformations.
The gauge transformations include the case when
$U$ is just a function (related to the exponentiation
of the center of $\gg$). These last transformations can be used to set
the component of $q(j)$ in the center of $\gg$, $q_c(j)$,
to any arbitrary
value; thus showing that it is not a dynamic degree of freedom but only
a purely gauge dependent quantity, so the hierarchy is completely
independent of the center of \gg.

A convenient choice
for the gauge slice has been proposed in~[\Ref{GEN3}], and we shall use
it in the following.
With this gauge choice, the integrable hierarchy
can be defined in terms of some $\Theta\,\in \,U_{-}(\s)$
through the equations
$$
{\cal L}_j = {\partial\over \partial t_j} + \Theta \left(
{\partial\over \partial t_j} - b_j\right) \Theta^{-1}\/,
\nfr{Laxgood}
and
$$
{\partial \Theta\over \partial t_j} = -\Neg\left(\Theta b_j\Theta^{-1}
\right) \Theta\/,\qquad j\in E\geq0\,.
\nfr{Hiertheta}
Therefore, by comparison with \Lax,
$$
q(j) = \Pos\left(\Theta b_j\Theta^{-1}\right) - b_j \in Q(j)\/.
\nfr{Qtheta}
Now, the zero curvature conditions \Zerocurv\
are just a consequence of the trivial identities
$$
\left[{\partial\over \partial t_i} -b_i ,{\partial\over \partial t_j}
-b_j \right] =0\/,\qquad i,j\in E\geq0\/.
\nfr{Zerotrivial}
It is worth noticing that our gauge choice {\it is
not} the standard Drinfel'd-Sokolov gauge; for example, for the original
KdV equation the two gauges correspond to
$$
q(1)\> =\> \pmatrix{\omega & 0 \cr \omega' -
\omega^2 & - \omega  \cr} \;\; {\rm and}\;\;
q_{DS}(1)\> =\> \pmatrix{0 & 0 \cr -u
& 0 \cr}\,,
\nfr{Exam}
where $'$ denotes a $x$-derivative, and $u=-2\omega'$.

The equations \Zerocurv\ can be interpreted as a system
of partial differential equations on some set of functions in a number of
different ways. For each {\it regular} element $b_k\in\ss$ ($k>0$), so
that $\gg$ admits the decomposition
$$
\gg = \ss \oplus {\rm Im}\>({\rm ad}\> b_k)\/,
\nfr{Regular}
we may regard $\Theta$ as a function of $q(k)$ and its
$t_k$-derivatives, and \Zerocurv\ becomes an integrable
hierarchy of partial differential equations on the function $q(k)$ (in
the language of~[\Ref{GEN1}], these are the ``type-I'' hierarchies of
equations). In this fashion, an integrable hierarchy of partial
differential equations is associated to the data $\{\gg,\ss,\s;b_k\}$.
Notice that there is a one-to-one correspondence between the solutions
of the two hierarchies associated to $\{\gg,\ss,\s;b_k\}$ and
$\{\gg,\ss,\s;b_j\}$; the maps being provided by \Qtheta. In this
sense, one does not distinguish between such hierarchies, even though
their hamiltonian structures can be very different~[\Ref{GEN2}].

\chapter{The additional symmetries.}

The derivation of the additional symmetries of the hierarchy follows
quite closely the arguments of~[\Ref{GEN1}] used to construct the flows
of the hierarchy. The idea is to obtain new operators
which commute with the Lax operators ${\cal L}_i$, but now allowing for
explicit
dependence on the $t_j$'s. In particular, using the semi-direct product
of the Virasoro algebra with $\gg$ (see the appendix), we shall find
an infinite number of operators commuting with the ${\cal L}_i$'s
that depend on the $t_j$'s. Our procedure is
quite similar to that of~[\Ref{ORLOV}] (see also~[\Ref{DICB}])
and~[\Ref{SEMI}]. As before, let us consider an affine  Kac--Moody
algebra \gg, a given Heisenberg subalgebra $\ss\subset\gg$ (whose
associated gradation is $\s'$), and an additional gradation
$\s\preceq\s'$. We shall need the following two lemmas.

\jump\noindent
{\bf Lemma 3.1.} {\it If $\,[\dd_{n}^{(\s)} + M ,{\cal
L}_j]=0$, where $M \in \gg$, and
$$
n\in{\Bbb Z} \geq \cases{-1,& if
$\s=\s_{\rm hom}\,$; \cr
0,& if $\s\succ\s_{\rm hom}\,$,\cr}
\nfr{Condition}
then
$$
\left[\dd_{n}^{(\s)} + \Pos(M), {\cal L}_j\right] =
- \left[\Neg(M), {\cal L}_j \right] \in Q(j)\/.
\nfr{Split}}

\sjump\noindent
{\it Proof.\/} The proof of this lemma proceeds by equating the grades of
the left- and right-hand sides of \Split, taking into account
that $[\dd_{n}^{(\s)}, \gg_0(\s)]=0$ and $[\dd_{n}^{(\s)},
\gg_{\geq0}(\s)]\subseteq \gg_{\geq nN_{\s} +1} (\s)$, as follows from
(A.13). With this in mind, the $\s'$-grade of the right-hand side is
$<j$ and the $\s$-grade of the
left-hand side is $\geq {\rm min}(0,nN_{\s} +1)$. Since both sides
must have the same grade, they actually lie in the intersection $Q(j)$ if
the value of $n$ is restricted by \Condition\ (recall that $N_\s = 1$
only if $\s=\s_{\rm hom}$; otherwise, $N_\s>1$). $\square$

\jump\noindent
{\bf Lemma 3.2.} {\it For any $n\in{\Bbb Z}$, let us define
$$
S_n\> =\> \Theta\, \widetilde{s}_n\, \Theta^{-1}\> - \> \dd_{n}^{(\s)}
\in \gg
\nfr{Sn}
where
$$
\eqalign{
\widetilde{s}_n \> =\>  \dd_{n}^{(\s')} &+ \sum_{j\in E>0} {j\over N_{\s'}}\,
t_j\, b_{j+nN_{\s'}} + {\alpha\over N_{\s'}}\, b_{nN_{\s'}} \cr
& + c\>\left(\left[ {1\over2\,N_{\s'}} \sum_{j+k+nN_{\s'}=0
\atop j,k\in E>0} jk\> t_j t_k +\,\vert n\vert \, \alpha\,
t_{\vert n\vert\, N_{\s'}}\right] \theta(-n) + \lambda\, \delta_{n,0}
\right)\/,\cr} \nfr{Sntilde}
where $\alpha$ and $\lambda$ are arbitrary constants ($\alpha$ is not
present if $0\notin E$), and $\theta(x)=1$
if $x>0$, otherwise vanishing (notice that  $S_n\in\gg$ because
$\dd_{n}^{(\s)} - \dd_{n}^{(\s')}\in \gg$, see (A.16)). Then:

\sjump
\itemitem{(i)} For any $n\in{\Bbb Z}$ and $j\in E\geq0$,
$$
[\dd_{n}^{(\s)} + S_n ,{\cal L}_j]=0\>,
\efr
\itemitem{(ii)} For any $m,n\in{\Bbb Z}$,
$$
\eqalign{
\left[ \widetilde{s}_m, \widetilde{s}_n\right] & = (m-n) \left(
\widetilde{s}_{m+n}\> -\> c\,\left[\lambda -{\alpha^2\over2\,
N_{\s'}}\right] \delta_{m+n,0}\right) \cr
& + {\widetilde{c}_v\over 12}\> m\,(m^2-1)\, \delta_{m+n,0}\/. \cr}
\nfr{Previra}}

\sjump\noindent
{\it Proof.} The proof of (i) follows from \Laxgood\ and the identity
$$
[\widetilde{s}_n, {\partial\over\partial t_j} - b_j]=0\/.
\nfr{Trivial}
The proof of (ii) is also straightforward by using (A.10),
(A.14), and
$$
[\dd_{m}^{(\s')}, b_j] = - {j\over N_{\s'}} b_{j + mN_{\s'}}\/.\ \square
\nfr{Semiheis}

\jump\noindent
We now define the following set of derivations
$$
\eqalign{
{\partial q(j) \over\partial \beta_n}\>& =\> -\left[\,\Neg(S_n)\,, \,
{\cal L}_j\,\right] \cr
& =\> +\left[\, \dd_{n}^{(\s)}\> + \>\Pos(S_n)\,, \,{\cal L}_j\,\right]\,,
\quad n\in{\Bbb Z}\geq\cases{-1,&
if $\s=\s_{\rm hom}\>$; \cr
0,& if $\s\succ\s_{\rm hom}\>$,\cr}}
\nfr{Newflows}
where the equality follows from Lemma 3.1. These derivations act on
$\Theta$ as
$$
{\partial \Theta \over\partial \beta_n} = \Neg(S_n)\>\Theta\/.
\nfr{Newflowtheta}
Obviously, as shown in Lemma 3.1, $\partial q(j)/\partial \beta_n \in
Q(j)$ for the values of $n$ indicated in \Newflows; moreover,
\Newflowtheta\ is consistent with our gauge choice $\Theta\in
U_-(\s)$.

The derivations in \Newflows\ actually define symmetries of the
hierarchy since, as we shall prove in the following proposition, they
commute with the flows of the hierarchy.

\jump\noindent
{\bf Proposition 3.3.} {\it The flows defined by \Zerocurv\ and
the derivations defined by \Newflows\ commute.}

\sjump\noindent
{\it Proof.} It will be sufficient to show that the flows and the
new derivations commute acting on a generic Lax operator. Therefore,
let us consider $i,j\in E\geq0$. Then,
$$
\eqalign{
& \left[
{\partial\over \partial t_i},
{\partial\over \partial \beta_n}\right] {\cal L}_j   = \left[\Neg
\left(\Theta b_i \Theta^{-1}\right) ,\left[ \Neg\left(
S_n \right) , {\cal L}_j\right] \right] \cr
&\quad + \left[ \Neg\left(\left[ \Neg\left( S_n\right) , \Theta b_i
\Theta^{-1} \right]\right), {\cal L}_j\right] \cr
&\quad
+\left[\Neg\left(\Theta{\partial \widetilde{s}_n\over\partial t_i}
\Theta^{-1}\right) , {\cal L}_j \right] -
\left[\Neg\left( S_n\right),\left[ \Neg\left( \Theta b_i \Theta^{-1}
\right) , {\cal L}_j\right] \right] \cr
&\quad
-\left[ \Neg\left(
\left[ \Neg\left( \Theta b_i\Theta^{-1}\right) , \Theta
\widetilde{s}_n \Theta^{-1}\right]\right), {\cal L}_j\right]\cr
& \qquad\qquad\qquad=-\left[ \Neg\left( \Theta \left[ {\partial\over
\partial t_i} - b_i,
\widetilde{s}_n \right] \Theta^{-1} \right), {\cal L}_j\right]\/,\cr}
\nfr{Commone}
which vanishes because of \Trivial. $\square$

\jump
Therefore, the new derivations \Newflows\ actually generate an infinite set
of additional symmetries of the integrable hierarchy. But they do not
provide additional flows because these new derivations do not commute
among themselves. Instead, as expected, they close on a subalgebra of the
Virasoro algebra.

\jump\noindent
{\bf Proposition 3.4.} {\it The derivations \Newflows\ have the
following commutation relations:
$$
\left[ {\partial \over\partial\beta_m}, {\partial \over\partial\beta_n}
\right] = (m-n){\partial \over\partial\beta_{m+n}}\/,
\nfr{Truncated}
for $m,n$ constrained as in \Newflows.}

\sjump\noindent
{\it Proof.} Again, it will be sufficient to consider the
commutator of two derivations acting on a generic Lax operator.
The result is
$$
\eqalign{& \left[
{\partial\over \partial \beta_m},
{\partial\over \partial \beta_n}\right] {\cal L}_j   = \cr
&\quad
+\left[\Neg\left( \left[\Neg\left(S_m\right) ,
\Theta\widetilde{s}_n\Theta^{-1}\right]
\right), {\cal L}_j \right) +
\left[\Neg\left( S_n\right) ,\left[\Neg\left( S_m\right) , {\cal L}_j
\right] \right]\cr
&\quad
-\left[\Neg\left( \left[\Neg\left(S_n\right) ,
\Theta\widetilde{s}_m\Theta^{-1} \right]
\right), {\cal L}_j \right) -
\left[\Neg\left( S_m\right) ,\left[\Neg\left( S_n\right) , {\cal L}_j
\right] \right]\cr
& \qquad\qquad\qquad =\left[ \Neg\left(\Theta\left[ \widetilde{s}_m,
\widetilde{s}_n\right]\Theta^{-1} -
\left[\dd_{m}^{(\s)}, \dd_{n}^{(\s)}\right] \right), {\cal L}_j
\right] \/,\cr}
\nfr{Commtwo}
which, using \Previra\ and (A.14), proves \Truncated. $\square$

\jump
Notice that the infinite set of additional
symmetries generated by the equations \Newflows\ and \Newflowtheta\
have been constructed in a completely representation independent fashion;
although it is important to remember that these expressions are only valid
in the special gauge chosen in section 2.

For the sake of illustration, we shall write the generators of the first
two additional symmetries in terms of the loop algebra representation of
\gg, ${\Bbb L}(g)\, = \,{\Bbb C}\,[z,z^{-1}]\otimes g$ (the central term
$c$ vanishes). In this
representation, $\dd_n = - z^{n+1} d/dz$, and $\dd_{n}^{(\s')}$ is given
by~(A.16). Let us consider a generalized hierarchy of the KdV type,
\ie, one for which $\s=\s_{\rm hom}$. Then,
$$
\eqalign{
& \Pos(S_{-1})\> =\> \sum_{j\in E>0} {j\over N_{\s'}}\, t_j \, \Pos\left(
\Theta\, b_{j-N_{\s'}}\, \Theta^{-1}\right)\,, \cr
& \Pos(S_0)\> =\> \sum_{j\in E>0} {j\over N_{\s'}}\, t_j \, \Pos\left(
\Theta\, b_{j}\,\Theta^{-1} \right)\> -\> {H_{\s'}\over N_{\s'}}\>
+\> {\alpha\over N_{\s'}}\,
b_0\,, \cr}
\nfr{SomeS}
where, as before, $\alpha$ is not present if $0\not\in I$. Then, if
$k<N_{\s'}$, it is straightforward to write the first two generators as
$$
\eqalign{
-{\partial q(k) \over \partial \beta_{-1}} \> &= \> \sum_{j\in E>N_{\s'}}
{j\over N_{\s'}}\, t_j \,{\partial \,q(k)\over \partial\, t_{j-N_{\s'}}}
\> -\> {d\over dz}\,(b_k + q(k))\cr
& \qquad -\> \sum_{0<j\in E\leq N_{\s'}} {j\over N_{\s'}} \left[\,t_j \,
P_{0[\s]}(b_{j-N_{\s'}})\>, \> {\cal L}_k\,\right]\,,\cr}
\nfr{Loopone}
and
$$
\eqalign{
-{\partial q(k) \over \partial \beta_0} \> &= \> -\>\left[\, z{d\over
dz}\,+\, {H_{\s'}\over N_{\s'}}\> , \> q(k)\,\right]\> +\> {k\over
N_{\s'}}\, q(k) \cr
&\qquad + \> \sum_{j\in E>0} {j\over N_{\s'}}\, t_j \,{\partial
q(k)\over \partial t_j}\> +\> {\alpha\over N_{\s'}}
\,{\partial q(k) \over \partial t_0}\,.\cr}
\nfr{Loopzero}
The first one generalizes the infinitesimal generator of the
Galilean transformation of the KdV equation, eq.\Galinf. In fact, in the
particular case of the KdV equation, \Loopone\ is
$$
-{\partial u\over \partial\beta_{-1}}\> =\> \sum_{j\in{\Bbb Z}>0}
\left(j\,+ {1\over2}\right)\, t_{2j+1}\, {\partial u\over \partial
t_{2j-1}}\> + \> 1\,,
\nfr{Galgen}
which, taking $t_j =0$ for $j>3$, is just \Galinf. Moreover, the result
that the transformation generated by $\beta_{-1}$ is a symmetry only of
hierarchies associated with $\s=\s_{\rm hom}$ is the generalization of
the well-known fact that the KdV equation is
Galilean invariant, whilst the mKdV equation (which has $\s=\s'$ being the
principal gradation) is not~[\Ref{DAS}].

In order to gain a better understanding of the transformation
generated by $\beta_0$, let us
specify the components of $q(k)$ with respect to a $\s'$-graded basis of
$Q(k)$: $q(k) =\sum_{r<k} q^r(k)\, e_r$, with $[d_{\s'},e_r]\,=\,r\,e_r$.
In terms of these components, \Loopzero\ is
$$
-{\partial q^r(k)\over \partial\beta_{0}}\> =\> \sum_{j\in E>0} {j\over
N_\s'}\, t_j\, {\partial q^r(k)\over \partial t_j}\> + \> {k\, -\, r\over
N_{\s'}}
q^r(k)\> + \>{\alpha\over N_{\s'}}\,{\partial q^r(k)\over \partial t_0}\,,
\nfr{Scalgen}
which is the generator of a scaling transformation under which the
scaling dimension of $q^r(k)$
is $k-r$, and that of $t_j$ ($j\not=0$) is $j$. This scaling symmetry of
the generalized hierarchies has been already discussed in~[\Ref{GEN2}].
In particular, \Scalgen\ generalizes \Scalinf.

\chapter{Additional symmetries and tau-functions.}

When \gg\ admits vertex-operator representations, {\it some} of
the integrable hierarchies defined by the equations
\Zerocurv\ can be described using the tau-function formalism [GEN3].
In terms of the tau-functions, the hierarchy consists
of an infinite set of bilinear equations known as Hirota equations, and
it is related to one of the integrable hierarchies constructed by Kac
and Wakimoto~[\Ref{KWI}]. Consequently, the additional symmetries
generated by \Newflows\ and \Newflowtheta\ can also be written as
transformations of the corresponding tau-functions.

\section{The tau-function formalism.}

For the sake of completeness, let us briefly review the construction of
integrable hierarchies within the tau-function formalism and their
connection with the zero-curvature hierarchies (for more
details see~[\Ref{KWI},\Ref{GEN3},\Ref{VOS}], or, in general, [\Ref{JAPOS}]).
The tau-function $\tau_{\s}$ associated to
an integrable highest weight representation $L(\s)$ of an affine
Kac--Moody algebra \gg\ is characterized by saying that it lies in the
$G$-orbit of the highest weight vector $v_\s$, with $G$ being the group
associated to \gg.

Let $\{u_i\}$ and $\{u^i\}$ be dual bases of the larger algebra
$\gg\oplus {\Bbb C}\,/d$, with respect to the non-degenerate bi-linear
form $(\cdot\mid\cdot)$. It can be shown~[\Ref{KWI},\Ref{PETK}]
that $\tau_\s$ lies in the $G$-orbit of $v_\s$ if and only if
$$
\sum u_i\otimes u^i (\tau_\s \otimes \tau_\s)\> =\> (\Lambda_\s\mid
\Lambda_\s)\, \tau_\s\otimes\tau_\s\/,
\nfr{Orbit}
where $\Lambda_\s$ is the eigenvalue of $\gg_0(\s)$ acting on $v_\s$.
Furthermore, the condition \Orbit\ is also equivalent to the statement
that $\tau_\s\otimes \tau_\s\in L(2\s)$.
It follows from the definition of the action of a group on a tensor
product that, for the representation $L(\s)$,
$$
\tau_\s =\bigotimes_{i=0}^{r} \left\{ \tau_{i}^{\otimes s_i}
\right\}\/,
\nfr{Taucomp}
where $\tau_i$ is the tau-function corresponding to the fundamental
representation with $s_j=\delta_{j,i}$.

When the representation $L(\s)$ is a vertex-operator representation,
\Orbit\ can be interpreted as a set of differential equations on the
tau-functions. In fact, they are precisely the Hirota equations of an
integrable hierarchy. Let us restrict ourselves to cases where
$\gg$ is the untwisted affinization of a simply-laced algebra (\ie,
$g$ is of $A$, $D$ or $E$ type).
In that case, level-one representations (or {\it basic} representations,
those for which $s_j=\delta_{j,i}$ for some $i$ such that $k_i=
k_{i}^{\vee}=1$) are isomorphic to the Fock space of any of the
Heisenberg subalgebras of $\gg$, which are classified by the conjugacy
classes of the Weyl group of $g$.

The Heisenberg subalgebra $\ss_w$, associated to some element of the Weyl
group (say $w$ up to conjugacy) is realized on the Fock-space
${\Bbb C}\, [x_j;\> j\in E>0]$ in the standard way:
$$
c=1\;\;  {\rm and}\;\; b_j=\cases{{\partial/\partial x_j}\,,& for
$j>0$\/,\cr \vert j\vert\, x_{\vert j\vert}\,,& for $j<0$\/.\cr}
\nfr{Fockheis}
A different treatment is required for the zero-graded elements of
$\ss_w$, which correspond to the invariant subspace of $w$. These
zero-modes are represented on the space
$$
{\Bbb C}\,(Q)=\{\exp (\beta\cdot x_0);\ \beta\in Q\}\/,
\nfr{Zeromodes}
where $Q$ is the root lattice of $g$ projected onto the invariant
subspace of $w$; $b_0$ acts as $\partial/\partial x_0$.

The level-one representation is isomorphic to ${\Bbb C}\,[x_j]\otimes
{\cal V}$, where ${\cal V}={\Bbb C}(Q) \otimes V$ is the zero-mode
space. Here, $V$ is an additional finite-dimensional vector
space~[\Ref{KPET},\Ref{VO}],
which is trivial (${\rm dim}(V)=1$) for the cases relevant to our
discussion~[\Ref{GEN3}]. The elements of $\gg$ not in $\ss_w$ are the modes
of vertex operators, and the derivation $d_{\s_w}$ is related to
the zero-mode of the Sugawara current.

Summarizing, the vertex operator representation of $L(\s)$ is realized
on the tensor product of fundamental representations, where $s_i$ gives
the multiplicity of the $i^{\rm th}$ fundamental representation in the
product (so any non-zero $s_i$ corresponds to $k_i=k_{i}^{\vee}=1$).
They will be carried by a tensor product of the Fock spaces:
$$
\bigotimes_{i=1}^{N_\s}\, \{\> {\Bbb C}\,[x_{j}^{(i)},\, j\in E>0]\,
\otimes\, {\cal V}\>\}\,,
\nfr{Tensor}
where $x_{j}^{(i)}$ indicates the Fock space variables of the
$i^{\rm th}$ space in the tensor product and $x_j\equiv
\sum_{i=1}^{N_\s} x_{j}^{(i)}$.

In the tau-function formalism of~[\Ref{KWI}] a hierarchy of Hirota
equations is associated to $\{g,w,\s\}$, \ie, a simply-laced finite
Lie algebra $g$, an element $w$ of the Weyl group of $g$ (up to conjugacy),
and a vertex operator realization of $L(\s)$, where $s_i=0$ if $k_i
\not=1$. Finally, the connection between the zero-curvature and the
tau-function formalism is established by the following theorem.

\jump\noindent
{\bf Theorem 4.1.} (Theorem 5.1 of [\Ref{GEN3}])
{\it There exists a map from solutions of the
Kac-Wakimoto hierarchy associated to the data $\{g,w,\s\}$ (with the
gradation associated to the Heisenberg subalgebra $\ss_w$ satisfying
$\s\preceq\s_w$, and also $s_i\not=0$ only if $k_i=1$) and a
zero-curvature hierarchy associated to the data $\{g^{(1)},\ss_w,\s;
b_k\}$, given by
$$
\Theta^{-1}\cdot v_\s ={\tau_\s (x + t)\over \tau_{\s}^{(0)}(t)}\/,
\nfr{Connection}
where $\Theta\in U_-(\s)$ gives $q(k)$ via \Qtheta, and $\tau_{\s}^{(0)}(x)$
is the $x_0$-independent component of $\tau$, \ie, the component
corresponding to $\beta=0$ in \Zeromodes.} $\square$

\jump
Notice that not all the zero-curvature hierarchies can be related to
tau-functions ($g$ has to be simply-laced, and
\s\ must correspond to products of level-one
representations). Conversely, not all the Kac-Wakimoto hierarchies
can be related to zero-curvature hierarchies because of the condition
$\s\preceq\s_w$.

\section{Additional symmetries of the tau-functions.}

In a vertex operator representation the generators of the Virasoro
algebra can be realized in terms of the elements of the Heisenberg
subalgebra through the Sugawara construction (see for example [\Ref{GO}])
$$
\eqalign{
& \dd_{n}^{(\s')}\mapsto L_{n}^{(\s')} = {1\over2\, N_{\s'}}
\sum_{i+j=nN_{\s'}} :b_i b_j:\> +\> \eta_{\s'}\>
\delta_{n,0}\,,\cr & \eta_{\s'}={1\over 4\, N_{\s'}^{2}}\>\sum_{j\in
I}j\, (N_{\s'}-j)\,,\cr} \nfr{Sugawara}
where $\,:\ :\,$ indicates that the product of elements of $\ss$
is ``normal-ordered''.
Consequently, acting on the Fock space, the generators $L_{n}^{(\s')}$
are second order differential operators; it will be convenient to write
them as $L_{n}^{(\s')} \equiv L_{n}^{(\s')} (\{x_i\}, \{\partial_j\})$,
where $\partial_j\equiv\partial/ \partial x_j$.

Taking into account \Newflowtheta\ and \Connection, one can easily
derive the action of the derivations \Newflows\ on the tau-functions:
$$
\eqalign{
{\partial\Theta^{-1} \over \partial\beta_n}\cdot v_\s \>&=\>
-\,\Theta^{-1}\, \Neg\left( S_n \right) \cdot v_s\cr
&= \>{1\over \tau_{\s}^{(0)}(t)}\, {\partial \tau_\s(x+t) \over\partial
\beta_n}\> - \> {\tau_\s(x+t)\over
\left[\tau_{\s}^{(0)}(t)\right]^2}\, {\partial  \tau_{\s}^{(0)}(t)
\over\partial \beta_n}\/.\cr} \nfr{Derione}
Now, using \Sugawara, it is straightforward to see that
$$
\widetilde{s}_n\> \mapsto\> L_{n}^{(\s')}\left( \left\{x_i + t_i\right\},
\left\{\partial_j +\alpha\delta_{j,0}\right\}
\right) \> + \> \left(\lambda -{\alpha^2\over2\,N_{\s'}}
\right)\delta_{n,0}\/.
\nfr{Key}
Therefore,
$$
\eqalign{
& S_n \cdot v_\s = P_{\leq0}\left(S_n\right)\cdot v_\s =
\left( \Theta \widetilde{s}_n\Theta^{-1} -
\dd_{n}^{(\s)} \right) \cdot v_\s \cr
&\quad = {1\over \tau_{\s}^{(0)}(t)}\>
\Theta\left[ L_{n}^{(\s')}\left( \left\{x_i + t_i\right\},
\left\{\partial_j +\alpha\delta_{j,0}\right\}
\right) + \left(\lambda -{\alpha^2\over2\, N_{\s'}}
\right)\delta_{n,0}\right]
\tau_\s(x+t) \cr
&\qquad - \eta_{\s}\delta_{n,0}\,; \cr}
\nfr{Deritwo}
the $x$-independent piece corresponding to
$P_{0[\s]}\left( S_n\right) \cdot v_\s $. Now, the comparison of
\Derione\ and \Deritwo\ allows one to prove the following proposition.

\jump\noindent
{\bf Proposition 4.2.} {\it In the tau-function formalism, the additional
symmetries of the hierarchy are generated by the derivations
$$
{\partial \tau_\s(x)\over \partial \beta_n} = -\left( L_{n}^{(\s')}
\left(\left\{ x_i\right\}, \left\{ \partial_j + \alpha\delta_{j,0}
\right\}\right) + \mu\delta_{n,0}
\right) \tau_{\s}(x)\/,
\nfr{Newflowtau}
where
$$
n\in{\Bbb Z}\geq\cases{-1,& if $\s=\s_{\rm hom}$; \cr
0,& if $\s\succ\s_{\rm hom}$\/,\cr}
$$
and $\alpha$ is an arbitrary constant that is not present if
$0\not\in E$. In general, $\mu$ is also an arbitrary constant, but it
has to vanish if $\s=\s_{\rm hom}$, as required by the commutation
relation $[\,\partial/\partial \beta_1\,, \,\partial/\partial
\beta_{-1}\,]\,=\, 2\,\partial/\partial \beta_0$.} $\square$

\jump
As we have discussed above, the vertex operator
representation of $L(\s)$ is realized on the tensor product of
fundamental representations, and the tau-function is also a tensor product
of ``fundamental'' tau-functions,
\Taucomp. Consequently, the derivations
\Newflowtau\ act on these ``fundamental'' tau-functions as
$$
{\partial \tau_i\over \partial \beta_n} = -\left( L_{n}^{(\s')}
\left(\left\{ x_j\right\}, \left\{ \partial_j + \alpha_i\delta_{j,0}
\right\}\right) + \mu_i\delta_{n,0} \right)
\tau_i(x)\/,
\nfr{Finaltaui}
with
$$
n\in{\Bbb Z}\geq\cases{-1& if $s_j=\delta_{j,0}\,$,\cr
0& otherwise.\cr}
\nfr{Condtau}
Of course, there is one equation for each component $s_i\not=0$
(having $k_i=1$). Again, $\alpha_i$ and $\mu_i$ are arbitrary
constants in
general, but all the $\mu_i$'s have to vanish if $\s=\s_{\rm hom}$.
These results agree with those obtained
in~[\Ref{VOS}], where it has been proven that if $\tau_{\s}$ is a
solution of the Hirota equations \Orbit, so is $\tau_{\s}+
\epsilon L_n\tau_{\s}$, $\epsilon\ll1$.

Finally, let us check again in this formalism that the transformation
generated by \Finaltaui\
with $n=-1$ actually generalizes the Galilean
transformation of the KdV equation. For the KdV equation, $w$ is the
Coxeter element,
$\gg=A_{1}^{(1)}$, $\s_w=(1,1)$, $\s=(1,0)$, and there is only one (scalar)
tau-function $\tau$. Then,
$$
-{\partial \tau\over \partial \beta_{-1}} \>= \> \left( \sum_{j\in{\Bbb
Z}>0} \left( j +
{1\over2}\right)\, t_{2j+1}\, {\partial \over\partial t_{2j-1}} +
{t_{1}^{2}\over 4}\right)\,\tau\,.
\nfr{Easyvir}
Therefore, using eq.\TauKdV, one recovers the action of this derivation on the
 original
variable of the KdV equation \Galgen.

\chapter{Generalizing the String equation.}

As explained in the introduction, one of the main recent motivations to
study integrable hierarchies of partial differential equations is their
importance in the matrix-model formulation of
two-dimensional quantum and topological gravity.
Let us summarize how the
integrable structure arises in the well-known case of the multi-matrix
models with chain-like interactions, which describe the coupling of
minimal conformal matter to two-dimensional gravity. The matrix model
leads to an integral over the eigenvalues of the matrix,
$\lambda$, and the basic idea, due to Douglas [\Ref{DOUG}], is that it
is possible to make the double-scaling limit on $\lambda$ and
$\partial/\partial\lambda$ considered as operators acting on certain space
of functions (the ``orthogonal polynomials''). The
final result is that the insertion of the eigenvalue $\lambda$ is
represented by a differential operator,
$Q$, and that of $\partial/\partial\lambda$ by another operator, $P$,
both of them acting on certain function $\Psi$. Therefore, we have the
two equations
$$
L\Psi \equiv \left(\hat\lambda -
Q\right)\Psi=0\,,\qquad \left({\partial\over \partial \hat\lambda} -
P \right)\Psi =0\,,
\nfr{PQ}
where $\hat\lambda$ is the double-scaled eigenvalue. The important
result is that $L$ is precisely the Lax operator of the
relevant hierarchy; in this case, the Drinfel'd-Sokolov $A_n$-KdV hierarchy,
where $n$ is the number of matrices in the model.
In the original argument of Douglas, $\Psi$ is a scalar function,
$L$ is a differential operator of order $n$, and the hierarchy
appears in the scalar Lax formalism~(see~[\Ref{DICB}]). Nevertheless,
in the general case, and depending on the details of
the double-scaling limit of the orthogonal polynomials,
the hierarchy is given more directly in the so called
``matrix'' Lax formalism [\Ref{MOORE}];
in which case $L$ is just a Lax operator like those of eq.\Lax\ written
in terms of the loop algebra ${\Bbb L}(g) = {\Bbb C}\,[z,
z^{-1}]\otimes A_n$, with $z=\hat\lambda$. Finally,
the compatibility condition for the two differential equations \PQ,
$$
\left[\, {\partial\over \partial\hat\lambda} - P\,,\, L\, \right] =0\,,
\nfr{Seqmm}
is just the ``string equation''. The string equation determines
the relevant solution of the integrable hierarchy  describing the
matrix-model.

Given that the zero-curvature hierarchies of~[\Ref{GEN1}-\Ref{GEN3}]
generalize the Drinfel'd-Sokolov hierarchies and share the same
structure, it is very tempting to consider the possibility that some of
them could also describe interesting
physical systems coupled to quantum or topological gravity in a similar way.
For that to be the case, then, at the very least, the hierarchy must admit
a generalization of the string equation \Seqmm.

The possibility of imposing additional constraints of the form \Seqmm\ is
ensured by the existence of the additional symmetries~[\Ref{LA}].
Recall that, in the loop algebra representation,
$\dd_n=-z^{n+1}d/dz$; hence, the invariance of the Lax operator under the
infinitesimal generator \Newflows\ with $n=-1$,
$$
-{\partial {\cal L}_k \over\partial \beta_{-1}}\>=\>\left[\, {d\over dz}\>
+ \>\Pos(S_{-1})\,, \,{\cal L}_k\,\right]=0\,,
\nfr{Loopalg}
for some ${\cal L}_k\equiv L$, is precisely a generalization of
\Seqmm. Moreover, one can check that the condition
$$
\Neg\left(\, S_{-1}\, \right)\>= \>0
\nfr{Steq}
is compatible with the hierarchy in the sense that it is preserved by all
the flows. Obviously, it induces the constraint \Loopalg, and it is the
natural choice for the ``generalized string equation'' for any zero-curvature
hierarchy of the KdV type\note{If the hierarchy is not of the KdV type,
\ie, $\s\not=\s_{\rm hom}$, then the natural generalization of the string
equation would be $\Neg(S_{0})\,=\,0$, in agreement with the results of the
unitary matrix-models.}. Moreover, when the hierarchy can be written in
terms of tau-functions, the constraint \Steq\ translates into an
$L_{-1}$-constraint for the (unique) tau-function,
$$
L_{-1}^{(\s')}(\,\{x_i\}\,,\,\{\partial_j\,+\,
\alpha\delta_{j,0}\})\, \tau\> =\> 0\,,
\nfr{Lcone}
according to \Newflowtau.

It is well known that the string
equation, together with the recurrence relations of the relevant
hierarchy, induces an infinite set of Virasoro
constraints~[\Ref{VER},\Ref{FUK}].
In the generalized case we are discussing, the generalized string equation
\Steq\ also induces an infinite set of constraints. To prove this, we
restrict ourselves again to the loop algebra representation. The crucial
observation is that $S_n\,= \, z^{j}\,S_{n-j}$, for any $n,j\in{\Bbb Z}$.
Therefore, the generalized string equation \Steq\ implies the following
infinite tower of constraints
$$
\Neg\left(\, S_{n}\, \right)\>= \>0\,, \qquad n\in{\Bbb Z}\geq -1\,.
\nfr{Steqn}
Even though we have used the loop algebra representation to prove
\Steqn, the
representation independence of the zero-curvature hierarchies, and the
fact that the eqs.\Steq\ and \Steqn\ are explicitly independent of the
center, ensure that the result is completely general. On the tau-function,
\Steqn\ are just the Virasoro constraints as expected:
$$
L_{n}^{(\s')}(\,\{x_i\}\,,\,\{\partial_j\,+\,
\alpha\delta_{j,0}\})\, \tau\> =\> 0\,\qquad n\geq-1\,.
\nfr{Lcn}

For the original case of the KdV hierarchy, which describes ordinary
quantum gravity, the set of Virasoro constraints \Lcn\ are complete in
the sense that they are equivalent to imposing the string equation
\Lcone\ along with
the fact that $\tau$ is the tau-function of the hierarchy. For the
more general hierarchies this is not the case, and the generalized
string equation
plus the hierarchy is equivalent to the Virasoro constraints \Lcn, as
we have shown, plus some additional constraints.
It is thought that these additional constraints satisfy a subalgebra
of a ${\cal W}$-algebra. The main evidence for
this belief comes from the study of the KP hierarchy, which contains
the Drinfel'd-Sokolov $A_n$-hierarchies as reductions. In this
case, and using the Grassmannian approach, it has been proven that the
string equation of the KP hierarchy induces an infinite set of
constraints satisfying a subalgebra of the ${\cal W}_{1+\infty}$-algebra.
Moreover, the
algebra satisfied by the constraints reduces to a subalgebra of the classical
$\W_n$-algebra when the KP hierarchy is reduced to the Drinfel'd-Sokolov
$A_{n-1}$-hierarchy~[\Ref{FUK}]. Directly in terms of the Lax operator
approach, and
again considering reductions of the KP hierarchy, it has also been proven
that the string equation induces an infinite set of constraints spanning a
subalgebra of the classical $\W_3$-algebra in the case of the Drinfel'd-Sokolov
$A_2$-hierarchy~[\Ref{GOER}].

The generalized string
equation \Steq\ does imply that the quantities $\Neg(S_{m_1}\cdots
S_{m_n})=0$, for $\sum_{i=1}^nm_i\geq-n$; however, we have not managed
to write these equations as constraints directly on the tau-function
and show that satisfy a \W-algebra.
It is clear, though, that these additional constraints are not
related to additional symmetries of the hierarchy\note{See~[\Ref{VOS}]
for similar comments about additional constraints and additional
symmetries within the tau-function approach.}. In the absense of a
direct construction of the constraints on the tau-function we shall
limit ourselves to some observations.

First of all, let us point out that whatever the additional constraints
are, they have to be consistent with the Virasoro constraints \Lcn, and
so form a closed algebra with this subalgebra of the Virasoro algebra.
So the most natural guess is that they satisfy a subalgebra of the
\W-algebra associated to the Casimirs of the relevant finite
Lie algebra $g$, being realized in terms of the Heisenberg subalgebra $\ss$
through a generalized Sugawara construction; the generators being
differential operators $W_{n}^{(\s')}(\{x_i\},\{\partial_j\})$.
We shall now prove that these additional
constraints would be compatible with the hierarchy. Let us consider
$$
R_n \cdot v_{\s} = \Theta W_{n}^{(\s')} \left(\{\partial_i\},
\{x_j + t_j\} \right)\Theta^{-1} \cdot v_{\s} =0\,,
\nfr{Wconst}
which follows from a $\cal W$-constraint on the tau-function:
$$
W_{n}^{(\s')} \left( \{\partial_i\}, \{x_j\}\right)\cdot \tau_{\s}
=0\,.
\nfr{Wcongen}
The time evolution of \Wconst\ is
$$
\eqalign{
{\partial R_n \over \partial t_j} \cdot v_{\s} & =
\left( \Theta {\partial \over \partial t_j}
W_{n}^{(\s')} \left(\{\partial_i\}, \{x_j + t_j\} \right)
\Theta^{-1}\right. \cr
&\qquad\quad - \left.\left[ \Neg(\Theta b_j \Theta^{-1}) ,
\Theta W_{n}^{(\s')} \left( \{\partial_i\}, \{x_j\}\right)
\Theta^{-1} \right] \right) \cdot v_{\s}  \cr
& =  \left[ \Pos\left( \Theta b_j \Theta^{-1}\right) ,
R_n \right]\cdot v_{\s} =0\,, \cr}
\nfr{cuenta}
for any $n$,
where we have used that $b_j\, \equiv\,
\partial/\partial x_j$ for $j\in
E\geq0$, \Wconst, and the fact that $v_{\s}$ is
annihilated by $\gg_{>0[\s]}$, and an eigenvector of
$\gg_{0[\s]}$.
Obviously, the reason for the consistency of \Wconst\ with the
hierarchy is just the identity
$$
\left[ {\partial \over \partial t_j } - b_j ,
W_{n}^{(\s')}\left( \{\partial_i\}, \{x_j +t_j\}\right)\right] =0\,,
\efr
and, of course, we could add arbitrary constant elements of $\ss$.

In the absence of a proof, we conclude this section
by making the natural conjecture that, for the cases where there exists a
tau-function formalism,
the generalized string equation \Steq\ induces an infinite set
of constraints on the tau-function which
satisfy part of the \W-algebra corresponding to the
Casimirs of $g$, for which there is a tower of
generators for each exponent of $g$. This
conjecture can be taken as a starting point to investigate the
possibility that some generalized integrable hierarchies could describe
two-dimensional physical systems including quantum gravity~[\Ref{TOP}].

\jump\jump
\centerline{\ttaarr Acknowledgments}

\sjump\noindent
JLM wants to thank P.G.~Grinevich, S.~Kharchev, A.~Mironov, A.Y.~Orlov
and A.M.~Semikhatov for their useful comments.

\jump
\appendix{Appendix}

In this appendix we review some of the properties of untwisted affine
Kac--Moody algebras that we use in the main text. A complete
treatment of such algebras may be found in~[\Ref{KBOOK}] and
references therein (see also~[\Ref{GO}]).

An affine Kac--Moody algebra $\gg$ is defined by a
generalized Cartan matrix $a$ of dimension $r+1$ and rank r, and is
generated by $\{h_i,e_{i}^{\pm};\; i=0,\ldots,r\}$ subject to the
relations
$$
\eqalign{
[\, h_i\,, \,h_j\,]\>= \>0\/, \qquad\quad & [\,h_i\, ,\,e_{j}^{\pm}\,]
\> =\> \pm a_{ij}\, e_{j}^{\pm}\/, \cr
[\,e_{i}^{+} \,,\, e_{j}^{-}\,] \>=\> \delta_{ij}\, h_{i}\/,
\quad\quad &
\left(\,{\rm ad}\, e_{i}^{\pm}\,\right)^{1-a_{ij}} \left(\,
e_{j}^{\pm}\, \right) =0\/.\cr}
\nfr{Algebra}
If $\gg$ is untwisted, $\gg=g^{(1)}$, this infinite algebra can be
realized as the untwisted affinization of a simple finite Lie algebra
$g$, generated by $\{h_i,e_{i}^{\pm};\; i=1,\ldots,r\}$, whose
(invertible) Cartan
matrix is $\overline{a}_{ij}\> =\> a_{ij}$ for $i,j=1, \ldots,r$.

Because $a$ has dimension $r+1$ and rank $r$, it has one left and
one right null eigenvector given by the combinations
$$
\sum_{i=0}^{r} k_{i}^{\vee} a_{ij} = \sum_{i=0}^{r} a_{ji} k_i =0
\/,\qquad j=0,\ldots,r\/,
\nfr{Kaclabels}
where $k_i$ and $k_{i}^{\vee}$ are the {\it Kac labels} and the {\it
dual Kac labels} of $\gg$, respectively,
the minimal positive integer numbers that satisfy
\Kaclabels. Notice that $k_0 = k_{0}^{\vee}=1$ when $\gg=g^{(1)}$.

The vector space generated by the $h_i$, $\hh=\sum_{i=0}^{r}{\Bbb
C}\,h_i$, is a maximally commuting subalgebra of $\gg$ known as the {\it
Cartan subalgebra}. From \Kaclabels, it follows that
$\gg$ has a centre ${\Bbb C}\>c$ where
$$
c=\sum_{i=0}^{r} k_{i}^{\vee} h_i \in \hh\/,\qquad [c,\gg]=0\/.
\nfr{Centre}

With respect to $\hh$, $\gg$ has a {\it root space decomposition}
$$
\gg =\bigoplus_{\alpha\in \hh^{*}} \gg_{\alpha}\/,
\nfr{Roots}
where $\hh^*$ is the dual space to $\hh$, and $\gg_{\alpha} =\{
x\in\gg \mid [h,x] = \alpha(h) x$ for all $h\in\hh\}$ is the {\it root
space} attached to $\alpha$; obviously, $\gg_0 = \hh$. The set of
$\alpha\in\hh^{*}\not=0$ such that $\gg_\alpha$ is not empty is called
the {\it root system} of $\gg$, and will be denoted by $\Delta$. The
set of roots $\{\alpha_i,\, i=0,\ldots,r\}$ such that $\alpha_i(h_j)
=a_{ij}$ are called the {\it simple roots} of $\gg$, and
$e_{i}^{\pm}$ is the step operator corresponding to $\pm\alpha_i$.
The vector space spanned by the simple roots is called the {\it
root space} of $\gg$. From  \Kaclabels, it
follows that the root space of $\gg$ contains a {\it null root}
$$ \rho =
\sum_{i=0}^{r} k_i \alpha_i \in \Delta\/,\qquad \rho(h) =0\/,
\nfr{Nullroot}
for any $h\in\hh$. Consequently, $\gg_{\alpha}=\gg_{\alpha+n\rho}$,
for any $\alpha\in\Delta\cup\{0\}$ and $n\in{\Bbb Z}$. When
$\gg=g^{(1)}$, the set of roots $\alpha_i$, with $i=1,\ldots,r$, are
the simple roots of the finite Lie algebra $g$, whose root system is
$\overline{\Delta}$.

\sjump
A derivation $d_{\s}$, with $\s=(s_0,s_1,\ldots,s_r)$ being a set of
$r+1$ non-negative integers ($s_0\not=0$), induces a $\Bbb Z$-grading
on $\gg$ which we label $\s$:
$$
[d_{\s}, h_i]=0\/,\qquad [d_{\s},e_{i}^{\pm}] = \pm s_i e_{i}^{\pm}\/;
\qquad i=0,\ldots,r\/.
\nfr{Derivation}
Under $d_\s$, $\gg$ has the eigenspace decomposition
$$
\gg=\bigoplus_{j\in{\Bbb Z}} \gg_j(\s)\/.
\nfr{Gradation}
Two particularly important choices for $d_{\s}$ are
$\s=(1,0,\ldots,0)$ and  $\s=(1,1,
\ldots,1)$, which induce the {\it homogeneous} and {\it principal}
gradation, respectively.
We shall often use the notation like $\gg_{>k}(\s) = \oplus_{i>k}
\gg_i(\s)$. There exists a partial ordering on the set of gradations,
such that $\s'\succeq\s$ if $s_{i}'\not=0$ whenever $s_i\not=0$.
An important property in the construction of integrable
hierarchies is that if $\s'\succeq\s$ then $\gg_0(\s')\subseteq
\gg_0(\s)$,
$\gg_{>0}(\s')\subset\gg_{\geq0}(\s)$,
$\gg_{<0}(\s')\subset\gg_{\leq0}(\s)$,
$\gg_{>0}(\s)\subset\gg_{>0}(\s')$, and
$\gg_{<0}(\s)\subset\gg_{<0}(\s')$.

It is useful to consider the larger algebra $\gg'=\gg\oplus{\Bbb
C}\>d_{\s}$, formed by adjoining a derivation $d_{\s}$ with
$[d_{\s},d_{\s}]=0$. The important difference between
$\gg'=\gg\oplus{\Bbb C}\, d_{\s}$ and  $\gg$ is that the former has an
invariant symmetric {\it non-degenerate}  bi-linear form
$(\cdot\mid\cdot)$, whereas for the latter the analogous  inner
product is degenerate. When needed, we shall normalize $(\cdot\mid
\cdot)$ as in~[\Ref{KBOOK}].
With respect to the Cartan subalgebra of
$\gg'$, $\hh'=\hh \oplus{\Bbb C}\, d_{\s}$, the root system of $\gg'$
is just
$$ \Delta' = \{\alpha + n\delta;\,
\alpha\in\overline{\Delta},\/ n\in{\Bbb Z}\} \cup\{ n\delta;\,
n\in{\Bbb Z}\not=0\}\subset\hh^{'*}\/.
\nfr{Rootsystem}
Notice that \Derivation\ means that $\alpha_i(d_{\s})= s_i$, which
implies that $\rho(d_{\s})=\sum_{i=0}^{r} k_i s_i\equiv N_{\s}$. The
root spaces $\gg_{\alpha}'$ attached to $\alpha\in \Delta'$ have finite
dimension; in particular, $\gg_{\alpha+n\delta}'$ has dimension one
for any $\alpha\in\overline{\Delta}$ and $n\in{\Bbb Z}$, while
$\gg_{n\delta}'$ has dimension $r$ for any $n\in{\Bbb
Z}\not=0$.

Let $d$ be the derivation inducing the homogeneous gradation,
\ie, corresponding to $\s=(1,0,\ldots,0)$. Then, for any other choice
of $\s$, $$
d_{\s} = N_{\s} d + H_{\s} \,, \;\; {\rm with}\;\;
H_{\s}= \sum_{i,j=1}^{r} s_i\,
\overline{a}_{ij}^{-1}\, h_j \in \hh\/,
\nfr{Deriform}
where $H_{\s}$ satisfies $\alpha_i(H_{\s}) = s_i$ for $i=0,\ldots,r$
(recall that $k_0=k_{0}^{\vee}=1$, and that the Cartan
matrix of the  finite Lie algebra $g$ is invertible). Therefore,
$d_{\s}/N_{\s} -  d_{\s'}/N_{\s'} \in \hh$, and the algebras formed
by adjoining different  derivations to $\gg$ are equivalent.

\sjump
The central objects in our construction of integrable hierarchies will
be the {\it Heisenberg subalgebras} of
$\gg$~[\Ref{KBOOK},\Ref{KPET}], $\ss= {\Bbb
C}\,c + \sum_{j\in E} {\Bbb C}\,b_j$, where $E= I + {\Bbb Z}\/N$,
and $I$ is a set of $r$ integers $\geq0$ and $<N$, for an integer
$N$; the  algebra being
$$
[\, b_j\,, \,b_k\,] \>= \>c\, j\, \delta_{j+k,0}\/.
\nfr{Heisenberg}
For each Heisenberg subalgebra there is an associated gradation $\s'$
that counts the grades of the elements of $\ss$,
$$
[\,d_{\s'}\,, \, b_j\,]\>= \>j\, b_j\/;
\nfr{Heisgrad}
the integer $N$ is precisely $N=N_{\s'}$.
The non-equivalent Heisenberg subalgebras are classified
by the conjugacy classes of the Weyl group of the finite Lie algebra $g$. The
connection between a Weyl group element, say $w$ (up to conjugacy), and
the associated Heisenberg subalgebra $\ss_w$, is that there is a lift of
$w$ to $\gg=g^{(1)}$, denoted $\widehat{w}$, which acts on the Heisenberg
subalgebra as
$$
\widehat{w} (b_j) = {\rm exp} \left( j \,{2\pi i\over N}\right) b_j\,.
\nfr{Endo}
In this case, $N$ is the order of $w$ whose eigenvalues are
${\exp}(2\pi i j/N)$, for $j\in I$. Two particular elements of the Weyl
group are the identity, for which $\s'=(1,0,\ldots,0)$ and $N=1$, and the
Coxeter element, for which $\s'=(1,1,\ldots,1)$, $N$ is the Coxeter
number, and $I$ is the set of exponents of the algebra.

\sjump
We shall make use of the {\it semi-direct product} of the Virasoro
algebra with $\gg$ for arbitrary gradations
that can be summarized as follows~[\Ref{KWAKI}]. Let us choose
a basis $\{E_{n\delta}^{(i)};\; i=1,\ldots,r\}$ of the root space
$\gg_{n\delta}'$ for $n\in{\Bbb Z}\not=0$, and $E_{\alpha+m\delta}$
of $\gg_{\alpha+m\delta}'$ for $\alpha\in\overline{\Delta}$
and $m\in{\Bbb Z}$,
and define the following set of derivations labeled
by a gradation $\s$:
$$
\eqalign{
&[\dd_{m}^{(\s)}, E_{n\delta}^{(i)}] = -n E_{(m+n)\delta}^{(i)}\/,
\qquad  [\dd_{m}^{(\s)}, \hh]=0\/,\cr
&[\dd_{m}^{(\s)},E_{\pm\alpha_i + n\delta}]
= -\left(n\pm{s_i\over N_{\s}}\right) E_{\pm\alpha_i +
(m+n)\delta}\/; \quad i=1,\ldots,r\/,}
\nfr{Semidir}
and
$$
[\,\dd_{m}^{(\s)}\,,\,  \dd_{n}^{(\s)}]\> =\> (m-n)\, \dd_{m+n}^{(\s)}
\>+\> {\widetilde{c}_v\over12}\>
m\,(m^2-1)\> \delta_{m+n,0}\/.
\nfr{Viralg}
These  derivations span a Virasoro algebra
$$
\Vir = \bigoplus_{m\in{\Bbb Z}}\, {\Bbb C}\, \dd_{m}^{(\s)}\/,
\nfr{Virasoro}
and \Semidir, together with $[\widetilde{c}_v,\gg]=0$, define
the {\it semi-direct product} of $\Vir$ and $\gg$, sometimes denoted as
$\Vir\semdir\gg$.

Let $\{\dd_n\/;\,n\in{\Bbb Z}\}$ be the Virasoro generators labelled by
the homogeneous gradation. Then it is easy to prove that~[\Ref{KWAKI}]
$$
\dd_{n}^{(\s)}\> =\> \dd_{n}\> -\> {H_{\s}^{(n)}\over N_{\s}\>}\>
+\> c\, {(H_{\s}\mid
H_{\s}) \over 2 N_{\s}^2}\> \delta_{n,0}\/,
\nfr{Viraform}
where $H_{\s}^{(n)}$ is an element of $\gg_{n\delta}$ such that
$[H_{\s}^{(n)},E_{\pm\alpha_i + m\delta}]= \pm s_i E_{\pm\alpha_i +
(m+n)\delta}$; obviously, $H_{\s}^{(0)}=H_{\s}$, \Deriform.
Therefore, $\dd_{n}^{(\s)} -
\dd_{n}^{(\s')}\in \gg$. It follows from \Viraform\ and  \Deriform\ that
$\dd_0=-d$ and
$$
\dd_{0}^{(\s)}\> =\> -{d_{\s}\over N_{\s}}\>
+\> c\,{(H_{\s}\mid H_{\s}) \over 2 N_{\s}^2}\/.
\nfr{Relation}

\jump
Integrable
highest weight modules of $\gg$ are defined in terms of a highest
weight vector $v_{\bf s}$, labelled, again, by a gradation $\s$ of $\gg$.
The highest weight vector is annihilated by $\gg_{>0}({\bf s})$, so
$$
e_i^+\cdot v_\s=0\/,\qquad i=0,\ldots,r\/,
\efr
and is an eigenvector of $\gg_0({\bf s})$ with eigenvalues
$$\eqalign{
h_i\cdot v_{\bf s}&=s_iv_{\bf s}\/,\qquad i=0,\ldots,r \cr
e^-_i\cdot v_{\bf s}&=0\ \ {\rm for\ any}\ i=0,\ldots,r\ {\rm
with}\ s_i=0\cr d_{\bf s}\cdot v_{\bf s}&=0.\cr}
\efr
The eigenvalue of the centre $c$ on the representation $L({\bf
s})$ is known as the level $k$:
$$
c\cdot v_{\bf s}=\sum_{i=0}^rk_i^\vee h_i\cdot v_{\bf
s}=\left(\sum_{i=0}^rk_i^\vee s_i\right)v_{\bf s},
\efr
hence $k=\sum_{i=0}^rk_i^\vee s_i$. In particular, the integrable
highest weight
representations with $k=1$ are known as the {\it basic\/} representations.
We shall use the notation $v_i=v_\s$, where $s_j=\delta_{ij}$, for the
highest weight vectors of the fundamental representations. In a
highest weight representation, the generators of the Virasoro algebra
can be realized in terms of the elements of $\gg$ through the Sugawara
construction~[\Ref{GO}].

\sjump
Associated to each Kac--Moody algebra $\gg$, there is a group $G$
formed by exponentiating the action of $\gg$ (see [\Ref{PETK}] for details).
We denote by $U_{\pm}(\s)$ and $H(\s)$ the subgroups formed by
exponentiating the
subalgebras $\gg_{>0}(\s)$, $\gg_{<0}(\s)$ and $\gg_0(\s)$,
respectively. The group acts projectively on the representations.

\references
\beginref
\Rref{NEW}{A.~Newell, {\sl Solitons in mathematics and physics\/},
Society for Industrial and Applied Mathematics (Philadelphia, 1985).}
\Rref{DAS}{A.~Das, {\sl Integrable Models\/}, World Scientific Lecture
Notes in Physics, Vol. 30 (Singapore, 1989).}
\Rref{HIR}{R.~Hirota, {\sl Direct methods in soliton theory\/}, in {\it
Soliton\/} page 157, eds. R.K.~Bullough and P.S.~Caudrey (1980).}
\Rref{JAPOS}{M. Jimbo and T. Miwa,
Publ. RIMS, Kyoto Univ. {\bf19} (1983) 943\newline
E. Date, M. Jimbo, M. Kashiwara and T. Miwa,
Publ. RIMS, Kyoto Univ. {\bf18} (1982) 1077.}
\Rref{KWI}{V.G. Kac and M. Wakimoto, {\sl Exceptional hierarchies of
soliton equations\/}, Proceedings of Symposia in Pure Mathematics.
Vol 49 (1989) 191.}
\Rref{DS}{V.G. Drinfel'd and V.V. Sokolov,
J. Sov. Math. {\bf30}
(1985) 1975.}
\Rref{KBOOK}{V.G. Kac, {\sl Infinite dimensional Lie algebras\/},
$3^{\rm rd}$ edition. Cambridge University Press 1990.}
\Rref{KPET}{V.G. Kac and D.H. Peterson, {\sl 112 constructions of the
basic representation of the loop group of $E_8$\/}, in: Symp. on
Anomalies, geometry and topology, eds. W.A. Bardeen and A.R. White.
World Scientific, Singapore, 1985.}
\Rref{GEN1}{M.F. de Groot, T.J. Hollowood and J.L. Miramontes,
Commun. Math. Phys.
{\bf145} (1992) 57.}
\Rref{GEN2}{N.J. Burroughs, M.F. de Groot, T.J. Hollowood and J.L.
Miramontes, Commun. Math. Phys. {\bf 153} (1993)
187; Phys. Lett {\bf B277} (1992) 89.}
\Rref{GEN3}{T.J. Hollowood and J.L. Miramontes, {\sl Tau-Functions and
generalized integrable hierarchies\/}, CERN preprint CERN-TH-6594/92 to
appear in Commun. Math. Phys.}
\Rref{LA}{H.~La,
Modern Physics Letters {\bf A6} (1991) 573; Comm. Math. Phys. {\bf 140}
(1991) 569.}
\Rref{GRIN}{P.G.~Grinevich, {\sl Nonisospectral symmetries of the KdV
equation and the corresponding symmetries of the Whitham equations}, to
appear in: Proceedings of the NATO ARW Singular Limit of Dispersion
Waves, (Lyon, France, 1991).}
\Rref{IBRA}{N.Kh. Ibragimov and A.B.~Shabat, Sov.~J. Phys. Dokl. {\bf
244} (1979) 57.}
\Rref{CHEN}{H.H. Chen, Y.C. Lee and J-E.~Lin, Physica~{\bf 9D}
(1983) 439\newline
P.G.~Grinevich and A.Y.~Orlov, {\sl Flag spaces in KP theory and Virasoro
action on ${\rm Det}\ \overline{\partial}_j$ and
Segal-Wilson tau-function\/},
in Problems of modern quantum field theory, eds. A.A.~Belavin,
A.U.~Klimyk  and A.B.~Zamolodchikov. Springer-Verlag, 1989.}
\Rref{CHINOS}{C.~Dengyan and Z.~Hangwei, J.~Phys. {\bf A24} (1991)
377.}
\Rref{ORLOV}{A.Y.~Orlov and E.I.~Shulman, Lett. in
Mathematical Physics {\bf 12} (1986) 171.}
\Rref{REV}{R. Dijkgraaf, {\sl Intersection Theory, Integrable Hierarchies
and Topological Field Theory\/}, IAS preprint IASSNS-HEP-91/91,
{\tt hep-th/9201003}, to appear in: Proceedings of the Carg\`ese
Summer School on New Symmetry Principles in Quantum Field Theory,
(1991)\newline
P.~Di Francesco, P.~Ginsparg and J.~Zinn-Justin, {\sl 2-D Gravity and
random matrices}, preprint SACLAY-SPHT-93-061, {\tt hep-th/9306153}.}
\Rref{VER}{R.~Dijkgraaf, H.~Verlinde and E.~Verlinde, Nucl. Phys.~B348
(1991) 435.}
\Rref{FUK}{M.~Fukuma, H.~Kawai and R.~Nakayama, Int. Jour. Mod.
Phys.~A6 (1991) 1385; Commun. Math. Phys.~143 (1992) 371.}
\Rref{WITT}{E. Witten, Surv. in Diff. Geom. {\bf 1} (1991) 243;
Nucl. Phys. {\bf B371} (1992) 191; \newline
M.~Kontsevich, Commun. Math. Phys.~147 (1992) 1.}
\Rref{SEMI}{A.M.~Semikhatov, {\sl Virasoro action and Virasoro
constraints on integrable hierarchies of the r-matrix type\/}, Lebedev
Phys. Inst. (Moscow) preprint, {\tt hep-th/9112016}; Nucl. Phys.
{\bf B366} (1991) 347.}
\Rref{DICK}{L.A.~Dickey, {\sl Additional symmetries of KP, Grassmannian,
and the string equation, I and II}, Univ. of Oklahoma preprints,
{\tt hep-th/9204092} and {\tt hep-th/9210155}.}
\Rref{VOS}{K. de Vos, Nucl. Phys. {\bf B375} (1992) 478.}
\Rref{KIEV}{T.J. Hollowood, J.L.~Miramontes and J.~S\'anchez Guill\'en,
{\sl Generalized integrability and two-dimensional gravitation},
preprint CERN-TH-6678/92, {\tt hep-th/9210066}, to appear in: Proceedings of
the
 Conference on
Modern Problems in Quantum Field Theory, Strings and Quantum Gravity
(Kiev, Ukraine, 1992).}
\Rref{DICB}{L.A.~Dickey, {\it Soliton equations and integrable systems},
World Scientific (1991).}
\Rref{PETK}{D.H.~Peterson and V.G.~Kac, Proc. Nat. Acad. Sci.
U.S.A.~80 (1983) 1778.}
\Rref{VO}{I.B. Frenkel and V.G. Kac, Invent. Math. {\bf62}
(1980) 23\newline
J. Lepowsky and R.L. Wilson, Commun. Math. Phys. {\bf62} (1978) 43\newline
V.G. Kac, D.A. Kazhdan, J. Lepowsky and R.L. Wilson, Adv. in
Math. {\bf42} (1981) 83\newline
J. Lepowsky, Proc. Natl.
Acad. Sci. USA {\bf82} (1985) 8295.}
\Rref{GO}{P.~Goddard and D.~Olive, Int. Jour. Mod. Phys. {\bf A1} (1986) 303.}
\Rref{DOUG}{M.~Douglas, Phys. Lett.~B238 (1990) 176.}
\Rref{MOORE}{C.~Crnkovi\'c and G.~Moore, Phys. Lett.~B247 (1991)
322\newline
G.~Moore, Commun. Math. Phys.~133 (1990) 261; Prog. Theor.
Phys. Suppl.~102 (1990) 255.}
\Rref{GOER}{J.~Goeree, Nucl. Phys.~B358 (1991) 737; \newline
S.~Panda and S.~Roy, {\sl `The Lax operator approach for the Virasoro and
the W-constraints in the generalized KdV hierarchy'}, ICTP preprint
IC-92-226, {\tt hep-th/9208065}.}
\Rref{TOP}{T.J.~Hollowood and J.L.~Miramontes, {\sl New topological
theories and conjugacy classes of the Weyl group}, preprint
CERN-TH-6758/92, {\tt hep-th/9212100}, to appear in Nucl. Phys.~{\bf B}.}
\Rref{KWAKI}{V.G.~Kac and M.~Wakimoto, Adv. in
Math. {\bf 70} (1988) 156.}
\endref
\ciao